\documentclass[floatfix,showkeys,twocolumn,showpacs,preprintnumbers]{revtex4}
\usepackage{array}
\usepackage{multirow}
\usepackage{reaction}
\usepackage{graphicx}
\usepackage{amssymb}

\begin{document}

\title{The effects of stochastic population dynamics on food web structure}

\author{Craig R. Powell}
\email{craig.powell@manchester.ac.uk}
\affiliation{Theoretical Physics Group, School of Physics and Astronomy, University of Manchester, Manchester, M13 9PL, UK}

\author{Richard P. Boland}
\email{richard.p.boland@postgrad.man.ac.uk}
\affiliation{Theoretical Physics Group, School of Physics and Astronomy, University of Manchester, Manchester, M13 9PL, UK}

\date{\today}

\begin{abstract}
We develop a stochastic, individual-based model for food
web simulation which in the large-population limit reduces to the
well-studied Webworld model, which has been used to successfully
construct model food webs with several realistic features.  We
demonstrate that an almost exact match is found between the population
dynamics in fixed food webs, and that the demographic fluctuations
have systematic effects when the new model is used to construct food
webs due to the presence of species with small populations.
\end{abstract}

\pacs{87.10.Mn, 87.23.Cc}

\keywords{ecological diversity; demographic stochasticity; ecological community model}

\maketitle

\section{Introduction}\label{Sec:Introduction}
There is a crucial distinction in biology between the proximal causes of
phenomena and the evolutionary explanation \citep{bul94}.  While the
former is useful in giving specific insight into a system being
studied, it is only through the understanding of the evolutionary
mechanisms in action that general principles can be understood.
Applying this understanding to ecosystems, the proximal explanation of
food web structure relates to the identity and habits of the species
within it \citep[for example][]{bla97,vat08}, but the underlying,
evolutionary explanation requires
understanding of the emergence and stability of those inter-specific
relations which contribute to the food web.  Whereas alternative models
of food web structure such as the niche model \citep{wil00} and the
cascade model \citep{sol98} are essentially static with respect to
their species composition, the Webworld model \citep{dro01} has been
used to demonstrate the formation of complex food webs stable to
evolutionary perturbation with several emergent features consistent
with observed ecosystems \citep{dro04,qui05a,qui05b}.  One aspect of
real ecosystems which it has
not been possible to reproduce in the Webworld model is the existence
of oscillations about a fixed point in the population dynamics.
Historic studies of population dynamics have tended to focus on
systems which experience oscillations, such as the Snowshoe Hare
\citep{vik08}, or Great Tit \citep{sae98}, but the Webworld model
shows monotonic approach towards a fixed point.  In part this is a
consequence of using population dynamics continuous in time, since
chaotic dynamics often emerge when large overshoots of the carrying
capacity are possible \citep{lan03}, an effect mitigated by using very
small time steps or effectively continuous dynamics.  \citet{kri06}
demonstrate the occurrence of limit cycles and chaotic behaviour even
using continuous population dynamics and a Holling type II functional
response.  It is likely that the explicit competition term in the
functional response used by Webworld prevents these cycles occurring.
It was, however, demonstrated by \citet{mck05a} that cyclic behaviour
could arise in a model using finite populations which disappears when
the infinite population limit is taken.  Since the continuous
population model of Webworld is equivalent to this limit, one
explanation for the absence of realistic population cycles is the
absence of individuals from the model, whose demographic stochasticity
could potentially lead to resonant oscillations in the population dynamics.

Stochasticity in ecosystem models tends to be divided into the three
categories described by \citet{lan03}.  The first of these,
demographic stochasticity, is simply the result of considering small
numbers of entities in a system.  For large populations, the law of
large numbers tends to ensure that birth and death processes,
happening at random to unrelated individuals, nevertheless average out
so that differential equations are an adequate description of the
processes.  For small populations it may be essential to the future of
a species how many, when, and in what order births and deaths occur.
In the case of most ecosystems, including those modelled by Webworld,
there exist at least some species with small populations \citep{mag03,SAD}.
If these species exert an influence on the ecosystem as a whole, for
instance by having a role as a top predator \citep{lyo05}, the
dynamics of the
system may crucially depend on the demographics of species with small
populations, and as a consequence demographic stochasticity cannot be
neglected even for the understanding of the broader ecosystem.  The
related effect of demographic heterogeneity, characterised by
\citet{mel08}, concerns the demographic stochasticity within species
of groups of heterogeneous individuals, for example males and females.
Even if the population of the species as a whole were constant, the
sex ratio is generally not strongly constrained by effects such as
prey abundance in the same manner as the population of the species as
a whole, and hence is more vulnerable to large-amplitude
fluctuations.  The Webworld model does not explicitly incorporate
sexual reproduction, or distinguish between individuals by sex, but in
Section~\ref{Sec:StrategyEffects} we show that demographic heterogeneity in terms of
foraging strategy can have important consequences for the food web.
The two further types of stochasticity discussed by \citet{lan03} are
not examined in this paper.  Of these, environmental stochasticity,
and indeed environmental periodicity, could be incorporated into the
Webworld model, but the work of \citet{mck05a} indicates that the
intrinsic fluctuations of demographic stochasticity are sufficient to
have large scale consequences on the population dynamics.  Because
demographic stochasticity can be naturally generated by the
probabilistic occurrence of events with characteristic frequencies, we
feel that it is more natural to introduce this as the first form of
stochastic influence on food web structure than to apply an external
forcing, which would require additional assumptions about the level at
which it acts (basal resource availability, death rate, etc.) and the
degree to which the effects on different species were correlated.  In
this paper we therefore develop an individual-based model designed to
replicate Webworld in the large-population limit.  The final common type
of stochasticity relates to measurement error.  As a measurement of a
real ecosystem will not yield an exact count of all individuals of
each species, the dynamics of the real system will differ from those
that would occur if the measurements were correct.  As such, 
even measurements of a constant population will show apparent
fluctuations.  Since the Webworld
mathematical model utilises exact populations, the population records
are at any moment in time precise, and this form of stochasticity is
not relevant in this case.

A second aim of this work is to introduce a model to which the results
previously published for Webworld are an approximation, valid under
conditions which need to be properly understood.  
It is more difficult to obtain precise results for the more detailed
model developed in this paper than for the parent Webworld model,
either analytically or numerically.  The parent model remains the most
appropriate level of detail at which to derive gross properties of
large ecosystems with relative computational economy.  In addition,
food webs established in the parent model can be used to grow food
webs adapted to stochastic population dynamics.  Imported food webs
can either be used as the basis for on-going evolution, or species can
be drawn from a parent-model `mainland' to construct immigrant
communities, as is done in the present paper.
%
%As is generally the
%case, the more detailed model developed in this paper is less easy to
%determine precise results for, either analytically or numerically.
%The virtue of the continuous Webworld model is therefore that it can
%be used to derive gross properties of large ecosystems with relative
%computational economy, and appropriately applied it can be used to
%establish complex ecosystems from which a more detailed model can
%derive more realistic systems either by continued evolution or, as
%examined in this paper, by a process of immigration.
%
By working towards a model in which individual actions can be
described, the consequences of differing plausible behaviours can be
examined.  In particular, an individual-based model such as that
introduced in this paper is the appropriate level of detail at which
to examine foraging strategy, or can be considered as an essential
step in deriving an agent-based model in which experience-derived
strategy optimisation, and life-history effects, can be examined.
The approach taken in this paper is to model species by discrete
individuals, who are affected by events such as death or the slow
digestion of consumed prey, and who interact in a pair-wise manner
between predators and prey.  This approach has the virtue that it can
be formally described in the same manner as a set of chemical
reactions, and the Gillespie algorithm \citep{gil77} can be used to
simulate essentially exact and typical histories.  In
Section~\ref{Sec:Webworld} we expand on the nature of this description;
in Section~\ref{Sec:ParentModel} we recap those aspects of the Webworld
model, more fully expounded in \citet{dro01}, which are necessary to derive
the individual-based model.  This derivation is presented in
Sections~\ref{Sec:Target} to \ref{Sec:ModelCompetitionForaging}.  In
Section~\ref{Sec:ModelVariants} we describe variants on the basic model
relating to foraging strategy selection in order to investigate the
sensitivity of the Webworld model to assumptions about decision making
which cannot be adequately investigated using continuous population
dynamics.  A summary of the stochastic model is given in
Section~\ref{Sec:ModelSummary}.

\section{Model}
\label{Sec:Webworld}
The Webworld model from which results have been published in previous
papers \citep{dro04,qui05a,qui05b,lug08a,lug08b} is stochastic to the extent that species are added as random
mutants of a randomly selected parent species, but the behaviour of
the system between speciation events is entirely deterministic.  In
particular, the population dynamics is represented by a set of
differential equations linking the rate of change of the population of
each species to its own population as well as the populations of its
prey, competitors and predators.  This is notably similar to the
representation of the dynamics of a chemical system in terms of
reaction rates.  Using the insight that such deterministic reaction
rate equations emerge from large numbers of interactions of discrete
molecules, and the desire to model ecosystem dynamics in terms of
individual organisms, in this paper we deduce the corresponding set of reactions
which gives rise to the Webworld model as its expected behaviour.  
Interactions changing the state of individuals will be written after
the manner of chemical reactions, which
follow the general model
\begin{reaction}
X_A+X_B\reactionarrow{\ensuremath{k}}{}X_C,
\label{Reac:Generic}
\end{reaction}
where $X_A$ and $X_B$ are individuals present before the interaction,
of types $A$ and $B$ respectively, and $X_C$ is the individual of type
$C$ which remains after the interaction.  Writing the abundance of these
types in turn as $N_A$ and $N_B$, the total rate at which
[\ref{Reac:Generic}] occurs is given by
\begin{equation}
  R=kN_AN_B,
\end{equation}
where $k$, written over the reaction arrow of [\ref{Reac:Generic}], is
the reaction rate coefficient.  Concrete examples of reactions are
given in Section~\ref{Sec:Target}.  In the
Webworld model the rate of predation is determined by a ratio-dependent
functional response, which we expect to be encoded in the appropriate
reaction rate coefficients.

Having chosen to make the stochastic Webworld model reaction-based, an
efficient means of implementing the dynamics in an essentially exact
manner is to use the Gillespie algorithm \citep{gil77}.  One of the assumptions we
retain in order to use this method is that spatial effects are not
relevant; in a chemical reaction scheme this is equivalent to
supposing the system to be well-mixed.  Under this assumption the
reaction rates can at all times be calculated in terms of the
abundance of the reactants without regard to their distribution.  The
original Webworld model makes similar assumptions about the absence of
spatial structure and hence is well-matched.  A model incorporating
spatial structure is likely to use spatial considerations as the basis
of the functional response underlying Webworld, and may therefore need
to describe in detail the individuals of each species and their
particular behaviour.  In addition to the direct interest in the
effects of demographic stochasticity on the assembly and functioning
of ecological communities, the model described in this paper can be
understood as a necessary step towards an agent-based model which
could explore the effects of individual-level behaviour and spatial
inhomogeneity.

For conciseness we will refer to the Webworld model with deterministic
population dynamics, discussed in \citet{dro01} and briefly introduced in
Section~\ref{Sec:ParentModel}, as the parent Webworld model, and the
model variant developed in this paper as the stochastic Webworld
model.  The term Webworld by itself refers to features generic to
both.

\subsection{Webworld}
\label{Sec:ParentModel}
The Webworld model consists of dynamics on three separated
time-scales, of which the longest corresponds to the introduction of
new species.  A detailed description of the parent model can be found
in \citet{dro01}.  In the present paper we do not consider the introduction of species
through evolution, but instead examine two types of system.  In the
first, an existing community, constructed in the parent model, is
interpreted in terms of discrete populations, and the evolution of the 
stochastic population dynamics is examined.  In the second type
of numerical experiment, we use a large community assembled in the
parent model as a species pool, from which individuals are drawn
during the construction of an immigrant community, in which the
population dynamics are again stochastic.  The subsequent discussion
of the parent model will only restate those features of the parent
model which are necessary to construct this non-evolutionary
stochastic model.

Interactions between species in the Webworld model are governed by two
evolved matrices, scores $S$ and competition $\alpha$.  Each pair of
species $i,j$ has an associated score $S_{ij}=-S_{ji}$ describing the
ability of one species to feed on the other.  A positive value of
$S_{ij}$ indicates that species $i$ is able to feed on species $j$,
larger values corresponding to preferred prey.  The anti-symmetric
nature of the matrix $S$ thus precludes mutual feeding and cannibalism.
The competition matrix, $\alpha$,
describes the degree to which species compete with one another when
feeding on the same food sources; this is described by a symmetric
matrix, which has a minimum value of 0.5 for highly distinct
species and a maximum value of 1 for intra-specific competition.  In
the parent Webworld model $S$ and $\alpha$ are derived by an intricate
system of evolution, but it has also been demonstrated \citep{Islands} that
varied communities can be assembled by immigration from an evolved
species pool.

The intermediate timescale of the Webworld model corresponds to the
population dynamics.  By encapsulating the per-capita income of
species $i$ from prey species $j$ in the functional response,
$g_{ij}$, the rate at which the population of each species $i$ changes
is described by
\begin{equation}
  \dot N_i=\lambda\sum_jg_{ij}N_i - \sum_jg_{ji}N_j - dN_i,
  \label{Eq:NDot}
\end{equation}
where the first term corresponds to the gain of resources through feeding,
the second term corresponds to losses to predation, and the final term
relates to the death of individuals through other processes.
Webworld equates energy resources to population, equivalent to an
assumption of equal body size for all individuals.  In these terms,
the ecological efficiency, $\lambda=0.1$, which appears in the first term of
(\ref{Eq:NDot}), corresponds to  the fraction of prey resources which can be
converted into new predator individuals.

The shortest time-scale of the Webworld model relates to the
adaptation of the functional response to changes in population.  Part
of this response is a direct consequence of the reduction in the prey
density, which reduces the rate at which individuals can be found.
The functional response also incorporates adaptation by foragers, who
select prey to maximize their income.  It is assumed that species as a
whole rapidly respond to changes in the relative abundance of
their prey and of their competitors.
Section~\ref{Sec:ModelVariants} discusses an alternative method we
consider by which the evolutionarily stable foraging strategy can be
identified, but an ecologically plausible mechanism is likely to
require an agent-based and spatial model, which will provide a
mechanism capable of reproducing the functional response.  The exact
form of the functional response as used by Webworld is
\begin{equation}
  g_{ij} = \frac{f_{ij}S_{ij}N_j}{bN_j+\sum_k\alpha_{ik}f_{kj}S_{kj}N_k},
  \label{Eq:g}
\end{equation}
which uses the scores, $S$, and competition, $\alpha$, already
described.  The term $b$ in the denominator limits population growth
for predator $i$ of species $j$ when $N_j\gg N_i$,  whereas the sum
limits the population of competing predators when their population is
non-negligible.  The term $f_{ij}$ represents the fraction of its time
that predator $i$ spends feeding on prey species $j$, and was
demonstrated by \citet{dro01} to have an evolutionarily stable strategy given by
\begin{equation}
  f_{ij} = \frac{g_{ij}}{\sum_kg_{ik}},
  \label{Eq:f}
\end{equation}
which can be found by iteratively solving (\ref{Eq:g}) and
(\ref{Eq:f}).

\subsection{Target reaction scheme}
\label{Sec:Target}
An essential modification made to the population dynamics in creating
this stochastic model is the division of each species into hungry
individuals, who are actively involved in feeding, and sated
individuals, who are not.  This standard assumption is used to derive
a functional response in which the predation rate for high prey
abundance has a limiting value \citep[for example][]{hui97}. An important consequence of this division
is that the rate of predation is not dependent on the total abundance
of the predator, but of its hungry members.  As prey become more
numerous, it should be expected that a smaller fraction of the
predator individuals remain hungry, and the predation rate per
predator individual becomes restricted.  This behaviour replicates an
essential feature of the Webworld functional response.

Having identified the need to divide species into hungry and sated
individuals, it follows from (\ref{Eq:NDot}) that four generic reactions
are required.  The first term in (\ref{Eq:NDot}) corresponds to
feeding, and hence to the reaction
\begin{reaction}
X_i^\prime+X_j\reactionarrow{\ensuremath{k_1}}{}X_i^*,
\label{Reac:Predation}
\end{reaction}
where $X_i^\prime$ and $X_i^*$ indicate an hungry and a sated individual
of species $i$ respectively, $X_j$ is any member of species $j$, and
reaction rate coefficient $k_1$ remains to be deduced.  To restore individuals to the
hungry state it is necessary to introduce the reaction
\begin{reaction}
  X_i^*\reactionarrow{\ensuremath{k_2}}{}X_i^\prime,
  \label{Reac:Hungering}
\end{reaction}
since no interaction is expected to be necessary for sated
individuals to become hungry.  To enforce the principle of energy
conservation not only on average, but in the details of the reaction
scheme, reproduction can be supposed to occur only after feeding, and
to be limited to one reproduction event per prey individual consumed.
This suggests an analogy to [\ref{Reac:Hungering}],
\begin{reaction}
  X_i^*\reactionarrow{\ensuremath{k_3}}{}2X_i^\prime,
  \label{Reac:Reproduction}
\end{reaction}
where a sated individual of species $i$ produces a hungry individual
while becoming hungry itself.  To complete the set of processes
associated with (\ref{Eq:NDot}) reactions corresponding to death are
required.  If death is equally likely to occur whether an individual
is hungry or sated,  the reaction can be written as
\begin{reaction}
  X_i\reactionarrow{\ensuremath{d}}{}\emptyset,
  \label{Reac:Death}
\end{reaction}
noting that, as with the prey individual in [\ref{Reac:Predation}],
[\ref{Reac:Death}] applies to both hungry and sated individuals.

\subsection{Minimal reaction scheme}
\label{Sec:Minimal}
To help deduce the reaction rate coefficients for the reactions in
Section~\ref{Sec:Target}, this section considers the simple scenario
of a single species, $i$, feeding on a resource of fixed abundance $R$.  The
expected rate of change of abundance of hungry individuals can then be
written
\begin{equation}
  \dot N_{i^\prime}=-k_1N_{i^\prime}R + k_2N_{i^*} + 2k_3N_{i^*} - dN_{i^\prime},
  \label{Eq:NDotPrime}
\end{equation}
where the four terms correspond to the four reactions in
Section~\ref{Sec:Target} in the order they appear. $N_{i^\prime}$ and
$N_{i^*}$ are the hungry and sated populations respectively, which sum
to give the total population, $N_i$. Correspondingly,
the number of sated individuals changes as
\begin{equation}
  \dot N_{i^*}=k_1N_{i^\prime}R - k_2N_{i^*} - k_3N_{i^*} - dN_{i^*},
  \label{Eq:NDotStar}
\end{equation}
with the population of the species as a whole changing according to
the sum of (\ref{Eq:NDotPrime}) and (\ref{Eq:NDotStar}),
\begin{equation}
  \dot N_i=k_3N_{i^*} - dN_i.
  \label{Eq:NDotPrimeStar}
\end{equation}
The simplified version of (\ref{Eq:NDot}) appropriate to this
scenario, using (\ref{Eq:g}) for the functional response,
is
\begin{equation}
  \dot N_i=\frac{\lambda S_{i0}R}{bR+S_{i0}N_i}N_i - dN_i,
  \label{Eq:NDotBasic}
\end{equation}
where the fixed resources have been identified as species 0, and $f=1$
occurs since no choice of prey is available.  As described in
Section~\ref{Sec:ParentModel}, $\alpha=1$ for all intra-specific
competition.

Progress can be made by finding an appropriate steady state by which
to eliminate time derivatives.  By assuming the population of full
individuals, $N_i^*$, to be a constant, (\ref{Eq:NDotStar}) becomes,
\begin{equation}
  N_{i^*} = \frac{k_1R}{k_1R+k_2+k_3+d}N_i,
  \label{Eq:NiStar}
\end{equation}
allowing us to eliminate from (\ref{Eq:NDotPrimeStar}) the population
of sated individuals.  This leaves
\begin{equation}
  \dot N_i=\frac{k_1k_3R}{k_1R+k_2+k_3+d}N_i - dN_i,
\end{equation}
which has obvious similarities to (\ref{Eq:NDotBasic}).  Assuming that
the $k$ values do not depend on $R$, it follows that
\begin{equation}
  k_3 = \frac{\lambda S_{i0}}b.
  \label{Eq:k3}
\end{equation}
Losses to the resources through
[\ref{Reac:Predation}] occur at rate
\begin{equation}
  k_1N_{i^\prime} R = \frac{k_2+k_3+d}{R+\frac{k_2+k_3+d}{k_1}}R,
  \label{Eq:rloss}
\end{equation}
and according to the Webworld model at rate $S_{i0}R/(bR+S_{i0}N_i)$. In the
limit of large R it follows that
\begin{equation}
  k_2 + k_3 = \frac{S_{i0}-bd}b.
  \label{Eq:k23}
\end{equation}
Because $S_{ij}$ is typically of order 10 for active feeding links whilst
$bd=1/200$, we neglect
the latter term in the numerator of (\ref{Eq:k23}) to obtain
\begin{equation}
  k_2=(1-\lambda) \frac{S_{i0}}b
  \label{Eq:k2}
\end{equation}
The fact that the rate coefficients of reactions
[\ref{Reac:Hungering}], [\ref{Reac:Reproduction}] and [\ref{Reac:Death}] do not depend on any
population implies that sated individuals can be considered
as non-interacting entities.  An implication of this in terms of predation is discussed
in Section~\ref{Sec:ModelSummary}.  Of all occurrences of reactions
[\ref{Reac:Hungering}] and [\ref{Reac:Reproduction}], a fraction
$\lambda$ result in reproduction, in line with the ecological
efficiency of the parent model.  The reaction rate of the predation
reaction is not so trivial, being given by
\begin{equation}
  k_1 = \frac1{N_{i}};
  \label{Eq:k1}
\end{equation}
intra-specific competition causes the feeding rate of each member of
the species to decrease in proportion to the total population, not
merely the hungry population who are concurrently foraging.
Section~\ref{Sec:ModelCompetitionForaging} examines how this rate
generalises to include inter-specific competition, and the effect of a
diversified foraging strategy in reducing the intra-specific
competition implied by (\ref{Eq:k1}).

\subsection{Foraging strategy and competition}
\label{Sec:ModelCompetitionForaging}
The generalisation of the reaction rate of predation must consider the
effects of intra- and inter-specific competition.  In the Webworld
model, species adopting a mixed foraging strategy increase their
reproduction rates by utilising more food sources, which is
mathematically described by the reduction in intra-specific
competition.  Multiple predator species feeding on one prey are taken
to be more effective at extracting resources from that prey species
than a single predator species due to differences in strategy, and
hence inter-specific competition is somewhat less detrimental than
intra-specific competition.  The effect of a mixed strategy by a
single predator species will be considered first.  The optimal
strategy, $f$, used by Webworld is the evolutionarily stable strategy (ESS);
an individual cannot increase its reproductive success by adopting any
other foraging strategy.  Given the implication of the analysis in
Section~\ref{Sec:Minimal} that the probability of reproduction by a
sated individual is $\lambda$, and hence independent of its food
source, it follows that the optimal strategy for a hungry individual
also corresponds to minimising the time before next becoming hungry.
In general this time depends on the prey species chosen by an
individual, and we write as $\tau_{ij}$ the expected time before a
hungry individual of species $i$, following a pure strategy of feeding
on species $j$, next becomes hungry.  For an optimal strategy
$\tau_{ij}$ is independent of $j$, and hence no individual can improve
its reproduction rate by changing strategy.  Importantly, this
condition is not the same as minimising the time before next feeding,
since the expected time in the sated state is
\begin{equation}
  \tau^*_{ij}=\frac{b}{S_{ij}},
  \label{Eq:tau}
\end{equation}
which is smaller for prey species corresponding to large $S_{ij}$.

Inspection of the Webworld functional response, (\ref{Eq:g}),
demonstrates that a single species following a mixed strategy
consisting of two prey has instantaneous population growth equivalent
to considering two populations following pure strategies, where the
sub-populations are given by $N_{ij}=f_{ij}N_i$.  In the absence of
both inter-specific competition and predation, and denoting the prey
as species 1 and 2, the total population growth is given by
\begin{eqnarray}
  \dot N_{i1} + \dot N_{i2}
  &=& \frac{\lambda S_{i1}N_1}{bN_1+S_{i1}N_{i1}}N_{i1} \nonumber \\
  + \frac{\lambda S_{i2}N_2}{bN_2+S_{i2}N_{i2}}N_{i2}
  &-& dN_{i1}
  - dN_{i2},
  \label{Eq:xyDot}
\end{eqnarray}
which is equivalent to treating components $N_{i1}$ and $N_{i2}$ as
separate species following pure strategies.  In terms of the
stochastic model, it follows that $N_{i1}$ is the total number of
individuals either foraging for prey 1 or sated having eaten that
prey, and hence we can write
\begin{equation}
  f_{ij}N_i = \phi_{ij}N_{i^\prime} + N_{i^*j},
  \label{Eq:phi}
\end{equation}
where $\phi$ is the foraging strategy of hungry individuals and
$N_{i^*j}$ is the number of individuals of species $i$ sated by having
consumed a member of species $j$.  The dependence of (\ref{Eq:phi}) on
$N_{i^*j}$ can be eliminated by considering the balance of reactions
affecting the population of sated individuals in the steady state,
where
\begin{equation}
  \frac{N_j}{f_{ij}N_i}\phi_{ij}N_{i^\prime} =
  \frac{S_{ij}}{b}N_{i^*j} + dN_{i^*j}.
  \label{Eq:Balance}
\end{equation}
The left hand side of (\ref{Eq:Balance}) represents the rate at which
foraging individuals become sated, the terms on the r.h.s being
the sum of hungering and reproduction, and death.  Using
(\ref{Eq:phi}) to eliminate $N_{i^*j}$ from (\ref{Eq:Balance}), the
foraging strategy of hungry individuals is found to be given by
\begin{equation}
  \phi_{ij}
  = \frac{N_i}{N_{i^\prime}}\left\{1+\frac{bN_j}{\left(S_{ij}+bd\right)f_{ij}N_i}\right\}^{-1}f_{ij}.
  \label{Eq:phiSolved}
\end{equation}
We note that, subject to the condition that $S_{ij}\gg bd$,
(\ref{Eq:phiSolved}) can be written as
\begin{equation}
  \phi_{ij}=\frac{f_{ij}S_{ij}N_j}{bN_j + f_{ij}S_{ij}N_i}\frac{N_i^2}{N_{i^\prime}N_j}f_{ij},
\end{equation}
where the first fraction is exactly the form of $g_{ij}$ in the parent
Webworld model.  Thus, in the absence of inter-specific competition,
the rate of feeding through [\ref{Reac:Predation}] is given by
\begin{equation}
  k_1N_{i^\prime}N_j=\frac{\phi_{ij}N_{i^\prime}N_j}{f_{ij}N_i}=g_{ij}N_i.
  \label{Eq:RFeeding}
\end{equation}
The appropriate foraging strategy for hungry individuals can be
calculated using (\ref{Eq:phiSolved}), and (\ref{Eq:RFeeding})
demonstrates that this results in dynamics consistent with the parent
model.

The reaction rate given by (\ref{Eq:RFeeding}) needs to be modified in
the presence of inter-specific competition.  By noting the argument
associated with (\ref{Eq:xyDot}), that species pursuing a mixed strategy
can without loss of generality be split into species of pure strategy,
it is possible to deduce the form of (\ref{Eq:RFeeding}) for
competition without strategy, and hence infer the behaviour of the
system when both competition and strategy are allowed.  First we
equate the population growth equations of the parent and stochastic
Webworld models for species $i$ feeding on prey species $j$,
\begin{equation}
  \frac{\lambda S_{ij}N_{i^*j}}{b} - dN_{ij} =
  \frac{\lambda S_{ij}N_jN_if_{ij}}{bN_j+\sum_k\alpha_{ik}f_{kj}S_{kj}N_k} - 
dN_{ij};
  \label{Eq:NDotEquated} 
\end{equation}
this can easily be simplified because the death terms and ecological
efficiency appear equivalently on both sides.  In steady state the
number of sated individuals, $N_{i^*j}$, can be found by setting the
population growth of strategy $i,j$ to zero, and hence
\begin{equation}
  k_{ij}\left(f_{ij}N_i - N_{i^*j}\right)N_j = \frac{S_{ij}+bd}{b}N_{i^*j},
  \label{Eq:kij}
\end{equation}
where $k_{ij}$ is the rate coefficient for the feeding reaction,
[\ref{Reac:Predation}], and the r.h.s.\,is the total rate at which the
number of sated individuals is reduced.  Rearranging (\ref{Eq:kij}) in
terms of the fraction $N_{i^*j}/f_{ij}N_i$ gives
\begin{equation}
  k_{ij}N_j = \frac{S_{ij}+bd}{b}\left\{\frac{f_{ij}N_i}{N_{i^*j}}-1\right\}^{-1}.
  \label{Eq:kijNj}
\end{equation}
This ratio can also be obtained from (\ref{Eq:NDotEquated}), from which
we obtain
\begin{equation}
  \frac{f_{ij}N_i}{N_{i^*j}} =
  \frac{bN_j+\sum_k\alpha_{ik}f_{kj}S_{kj}N_k}{bN_j}.
  \label{Eq:fNOverN} 
\end{equation}
Substituting (\ref{Eq:fNOverN}) into (\ref{Eq:kijNj}) gives
\begin{equation}
  k_{ij} = \frac{S_{ij}+bd}{\sum_k\alpha_{ik}f_{kj}S_{kj}N_k},
\end{equation}
which is similar in form to (\ref{Eq:g}).  If we again assume that
$S_{ij}\gg bd$, this can be written
\begin{equation}
  k_{ij} = \left\{\sum_k\alpha_{ik}f_{kj}\frac{S_{kj}}{S_{ij}}N_k\right\}^{-1},
  \label{Eq:kijSolved}
\end{equation}
and is clearly seen to simplify to (\ref{Eq:k1}) for a pure strategy
in the absence of inter-specific competition, where $\alpha_{ii}=1$.
For the case in which a single predator has reduced intra-specific
competition due to adoption of a mixed strategy, (\ref{Eq:kijSolved})
gives
\begin{equation}
  k_{ij} = \frac1{f_{ij}N_i},
\end{equation}
where the appropriate forager and prey populations are
$\phi_{ij}N_{i^\prime}$ and $N_j$ respectively, giving
(\ref{Eq:RFeeding}).

It is now possible to demonstrate that the ESS of the parent model
corresponds as expected to the optimal individual behaviour, which
minimises the time until next becoming hungry, when reproduction might
occur.  For species $i$ feeding on prey species $j$ this time is
\begin{equation}
  \tau_{ij}
  = \frac{\sum_k\alpha_{ik}f_{kj}S_{kj}N_k}{S_{ij}N_j}
  + \frac{b}{S_{ij}},
  \label{Eq:tauij}
\end{equation}
where the first term is the reciprocal rate of feeding, and the second
is the reciprocal of the total rate of hungering and reproduction.  Given the
definition of $g$ in (\ref{Eq:g}), it is clear that
\begin{equation}
  \tau_{ij} = \frac{f_{ij}}{g_{ij}}.
\end{equation}
For the ESS, the foraging strategy $f$ is related to the functional
response $g$ by
\begin{equation}
  f_{ij} = \frac{g_{ij}}{\sum_kg_{ik}},
\end{equation}
and hence the expected time for an individual of species $i$ to next
become hungry is given by
\begin{equation}
  \tau_{ij} = \left\{\sum_kg_{ik}\right\}^{-1},
\end{equation}
which is independent of prey as expected.

\subsection{Variants}
\label{Sec:ModelVariants}
Two basic variations of the stochastic Webworld model are suggested by
the details of the behaviour seen above, which relate to anti-predator
behaviour and to decision making.  Having shown that the rate
coefficients of [\ref{Reac:Predation}] and [\ref{Reac:Hungering}]
should be independent of population, it follows that sated individuals
are in no way influenced by the population of their own species or of
prey, and to some degree can be said to be isolated from the community
for the period of time they remain sated.  It is reasonable to assume
that for real animals subject to this condition, this period of time
would be best spent in a den or other refuge offering relative safety
from predation.  Regardless of the means by which such safety might be
effected, which is more properly the concern of an agent-based model
using spatial and individual-rationality considerations, we consider
it proper to examine whether the vigilance of sated individuals has
significant consequences for the model dynamics.  We therefore denote
as variant {\emph a} of the model the case in which sated individuals
are equally susceptible to predation as foraging individuals, and as
variant {\emph b} the case in which sated individuals cannot be predated
at all.  If it can be demonstrated that no significant differences
exist between these two cases, it is to be expected that reasonable
intermediate cases can also be unproblematically related to the parent
model.  In so far as it makes no difference to the dynamics, we prefer
to adopt variant {\emph b} as the default case, since this suggests a more
optimal individual behaviour.

In the parent model, foraging strategy is taken to adapt to the current
population of all species on the shortest possible time-scale, so that
at all times all individuals of all species follow the evolutionarily
stable strategy, which is computationally arrived at by considering
the effects on (\ref{Eq:g}) of prey availability and competitor
abundance.  A more plausible model of strategy selection would
incorporate at least the effects of limited knowledge, since
individual predators must make decisions based on observation or
communicated information rather than directly based on global
population data.  Such decision making is also likely to be based on
spatial distribution, and we therefore defer both informational and
spatial considerations to the agent-based model we foresee as the next
development of Webworld toward an ecosystem model grounded in
individual-level considerations.  For the purposes of this paper we
consider a method of strategy selection such that the eventual,
detailed behaviour model is likely to lie within the range
investigated here.  Variant I denotes the case in which each time the
population of any species changes, the strategy of each species is
updated, and all individuals follow the ESS.  For variant II we attempt
to recover the ESS through population dynamics.  In this case we
assign to each individual a prey species which is their only prey for
life, and are able to approximately track the ESS by causing offspring
to adopt the same foraging strategy as their parent.  Strategies which
are under-represented in the population of a species have a higher
reproduction rate than strategies which are over-represented, leading
to their replenishment and convergence toward the ESS.  A complication
is that the population of each strategy is subject to demographic
stochasticity, and hence at risk of extinction.  With probability
$\mu$ we assign each newborn individual to a strategy selected at random using
the contemporary ESS as a probability distribution.  This allows
extinct strategies to be repopulated without losing large numbers of
individuals to unviable strategies.  We investigate the results for
$\mu$ in the range $0.01\le\mu\le1$, and use $\mu=0.1$ as the standard
value.

\subsection{Summary}
\label{Sec:ModelSummary}
The implementation of the stochastic Webworld model requires that each
species, population $N_i$, be divided into sub-populations according
to the number of prey it possesses.  The number of individuals
actively foraging is given by $N_{i^\prime}$ while the sated
individuals are divided into groups according to the last prey item
consumed, with populations $N_{i^*j}$.  Sated individuals return to
the foraging population at a rate dependent on the prey species
consumed, returning more rapidly for prey species corresponding to a
large score, $S_{ij}$.  In variant I foraging effort is divided in the
same manner as in the parent model, according to the evolutionarily
stable strategy (ESS) given by (\ref{Eq:phiSolved}).  In the more
stochastic variant II each individual pursues a single prey species,
and the relative reproductive success of different foraging strategies
contributes to the identification of the ESS by natural selection.
For small forager populations it is necessary to avoid the permanent extinction
of strategies by allowing mutant offspring to follow a strategy
different from that of their parent, which occurs with probability
$\mu$.

The second variation in the stochastic model that we consider is the
ability to feed on sated individuals.  It is implicit in the
deterministic population dynamics that all individuals of a species
are equally vulnerable to predation, but as shown in
Section~\ref{Sec:Minimal}, sated individuals are not influenced by any
other individuals of their own or prey species.  Following such work
on optimal strategies as \citet{hou93}, it may be reasonable to assume that
individuals not engaged in foraging are able either to seek refuge
where they are not vulnerable to predators or to adopt some vigilance
strategy that removes the risk of predation at the expense of being
unable to simultaneously forage.  This being the case, and noting that
at the steady state the fraction of sated individuals is
\begin{equation}
  \frac{N_{i^*j}}{N_{ij}}=\frac{bN_j}{bN_j+\sum_k\alpha_{ik}f_{kj}S_{kj}N_k},
\end{equation}
which is typically of the order $10^{-2}$ for
$N_{\rm prey}/N_{\rm predator}\sim10$ and $S\sim10$, 
it is not strongly disruptive to the model to assume that only hungry
individuals are susceptible to predation.  Because we
believe it to have the most plausibility, and most potential for
developing a more detailed model, we use variant II\emph{b} as the default
model against which others are compared.

\section{Results by phenomenology}
\label{Sec:Phenomena}
In order to compare the newly developed stochastic model with the
parent model as closely as possible, we contrast their behaviour in
three sets of circumstances.  Each of these utilises the same
`mainland' community of species, evolved in the parent model and
containing 103 species, in order that incidental differences are
minimised.  The simplest comparison between the parent and stochastic
models is to directly compare the population dynamics, for which
purpose it is preferable to make use of a food web of fewer species and
lesser complexity than the mainland.  \citet{Islands} introduced the method of using an
evolved Webworld ecosystem as a species pool from which an `island'
community could be assembled by the addition of many species under the
action of the population dynamics.  Two small webs which resulted from
this process, which we refer to as Islands A and B, are drawn in
Table~\ref{Tab:Webs}, and were used to study the population dynamics in
the absence of the introduction of any further species.  These results
were repeated for several other webs of differing size as a test of
generality.  The labels associated with species indicate their trophic
level, so species P1 is a `plant', H1 a `herbivore', and C1 a
`carnivore'.  Species E is the environment, population $R$, which
provides a constant food source for all plants.

The second comparison of the two models is to create
communities directly in the stochastic model.  Because island creation
restricts the set of species to those present in the mainland
community, the results are more directly comparable than are
those for independently evolved communities, and the process of immigration
produces mature communities much more rapidly than is achieved by
evolution from a single ancestral species.  As such, the second test
of the similarity of the two models was performed by statistical
comparison of islands constructed under the two types of population
dynamics.  A third test, referred to as the `reduction' process, is
to take a large food web and record its behaviour as the resources
supplied to basal species, $R$, are reduced.  Using the mainland
community as the initial food web, reduction of $R$ tends to reduce
the population of each species by the same factor until one or more
species go extinct.  Because the food web structure changes at this
point, the subsequent behaviour of the food web is not trivially
predictable.  Under the population dynamics of the parent model
the reduction process
follows a deterministic trajectory, which can be compared to an
example trajectory derived from the stochastic model.

\begin{figure}
  \includegraphics[width=0.45\textwidth]{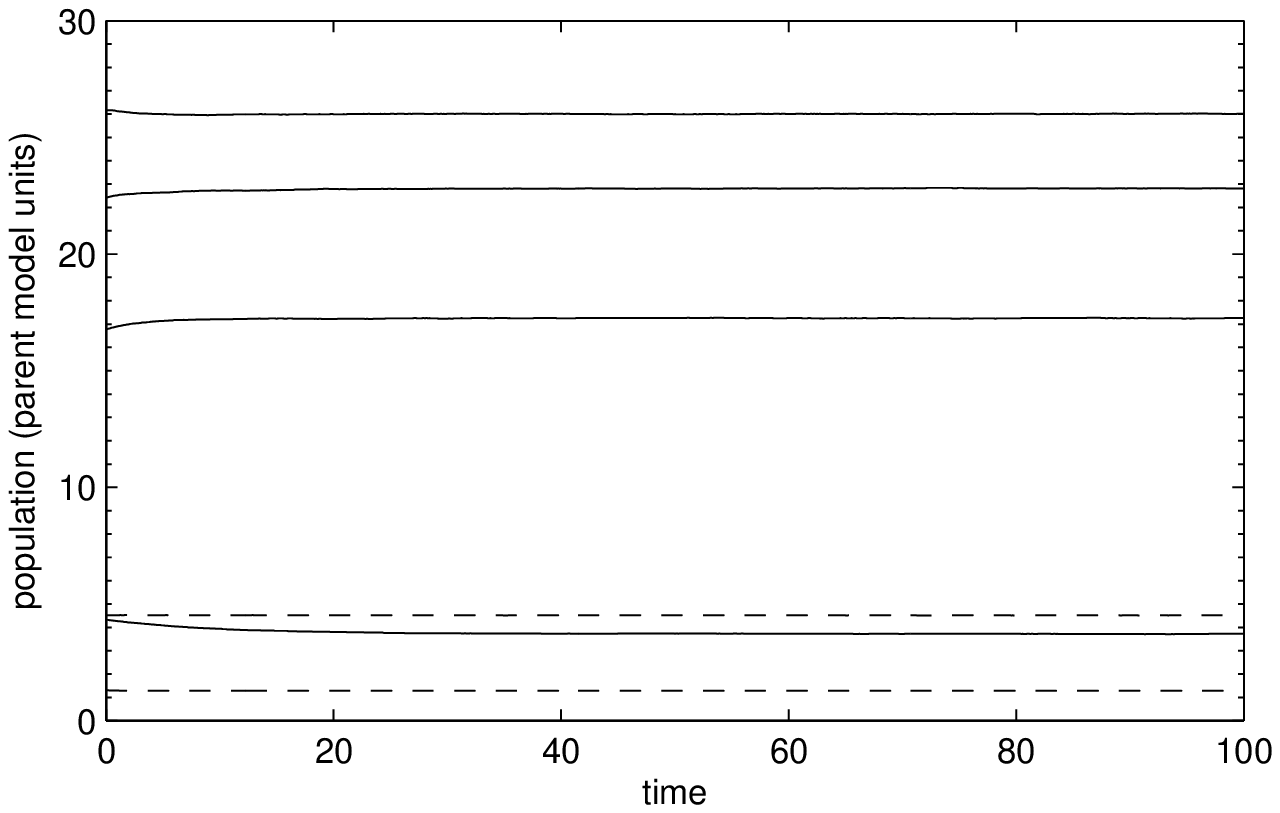}
  \caption{
    The ensemble mean of time series generated with the stochastic
    population dynamics for Island A, using variant II\emph{b} with
    $\mu=0.1$ and scaling $\Omega=300$.  Population sizes are divided
    by $\Omega$ for comparability.  Solid lines mark species
    on the lower trophic level (`plants'); dashed lines mark species on
    the upper trophic level (`herbivores').
  }
  \label{Fig:EnsembleTimeSeries300}
\end{figure}

\begin{figure}
  \includegraphics[width=0.45\textwidth]{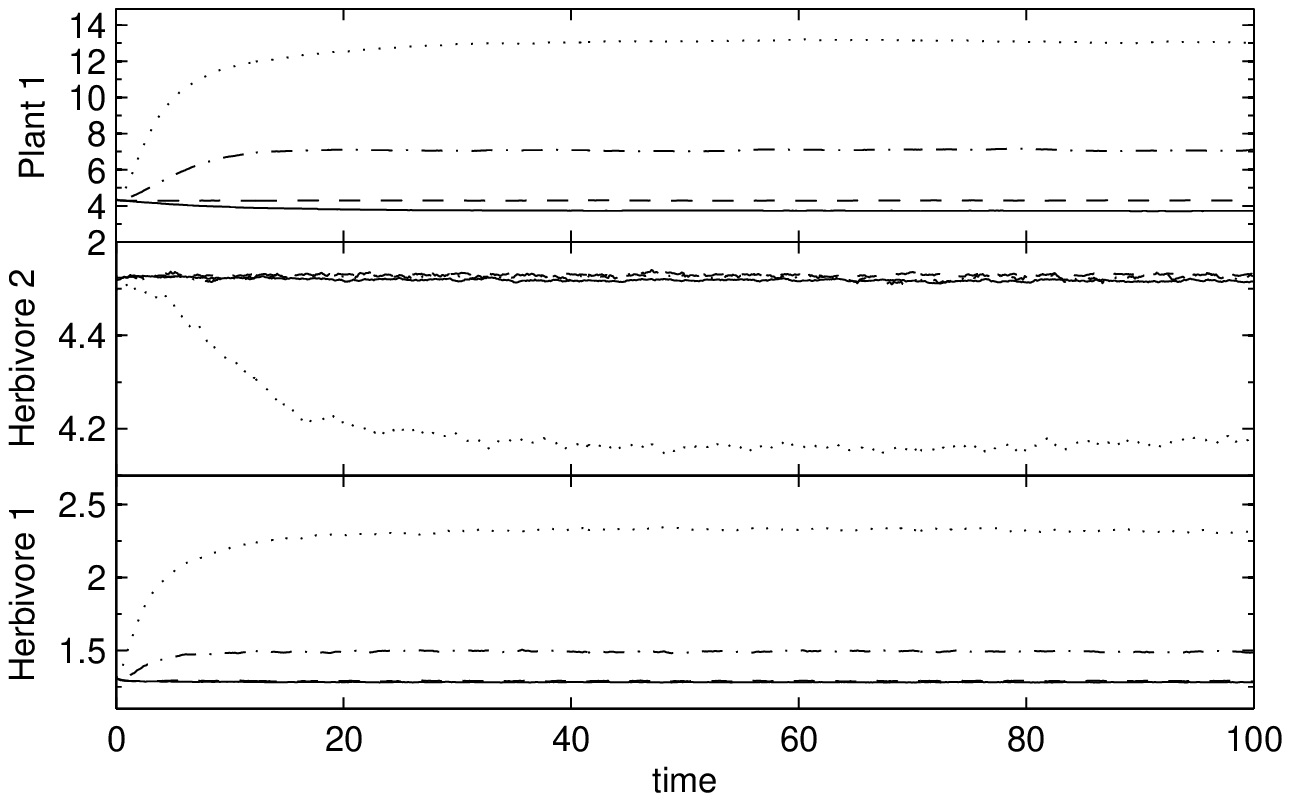}
  \caption{
    Selected species from Island A showing the effect of reducing
    the scaling from $\Omega=300$ (solid) through $\Omega=100$
    (dashed) and $\Omega=30$ (dot-dashed) to $\Omega=10$
    (dotted). Note that the $y$-axes are independently scaled and also
    overlap with each other.
  }
  \label{Fig:EnsembleScaling}
\end{figure}

\begin{table}
  \begin{tabular}{cc}
    \parbox[b]{.2\textwidth}{
      \begin{flushleft}
	Island A consists of two herbivore each feeding predominantly on pairs of
	plants. However, the most populous herbivore focuses some feeding attention
	on plant P1 and also directly on herbivore H1
      \end{flushleft}
    }
    & \parbox[b]{.25\textwidth}{\center\includegraphics[scale=0.3]{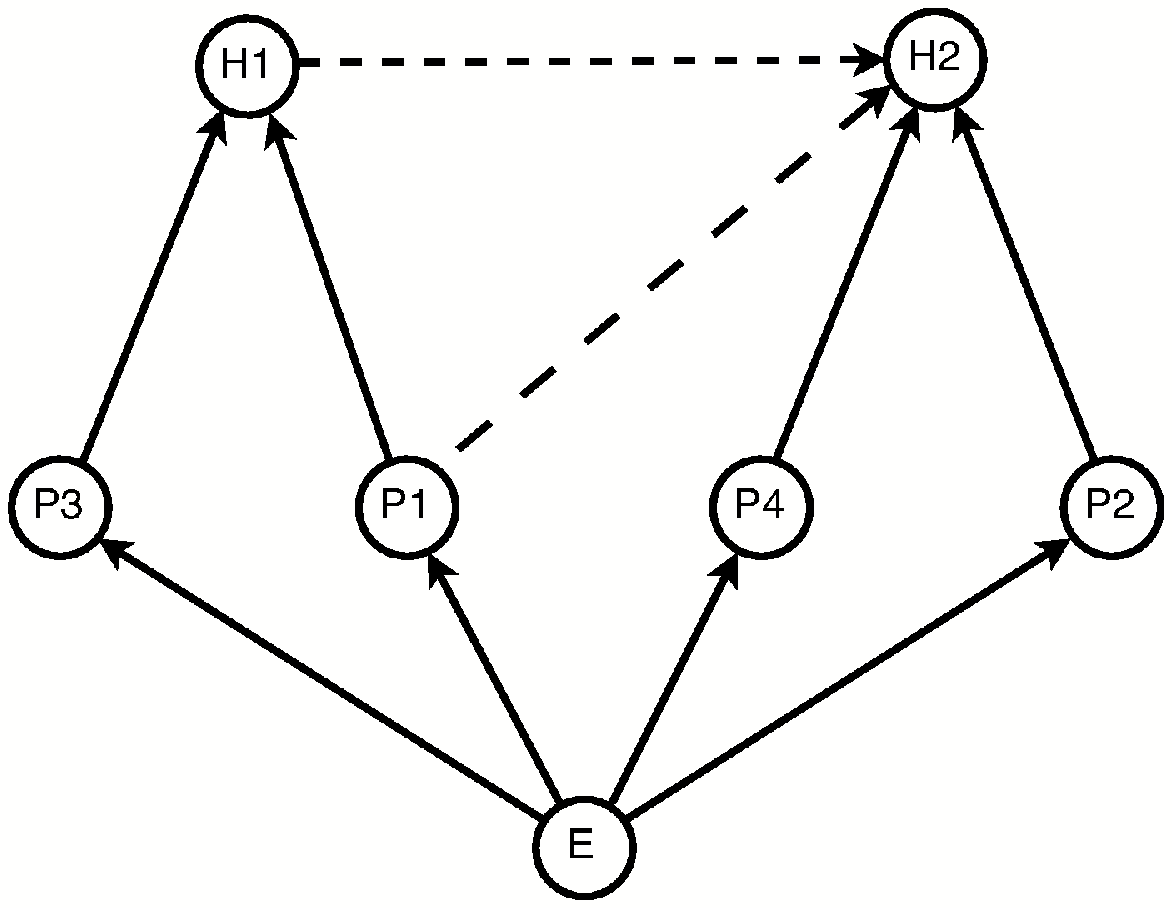}}\\ 
    \hline
    \parbox[b]{.2\textwidth}{
      \begin{flushleft}
	Island B is already a highly complex web with interesting adaption to
	increasing stochasticity. The omnivory of the carnivore and herbivore
	H2 are not seen in the parent model.\vspace{5mm}
      \end{flushleft}
    }
    & \parbox[b]{.25\textwidth}{\center\includegraphics[scale=0.3]{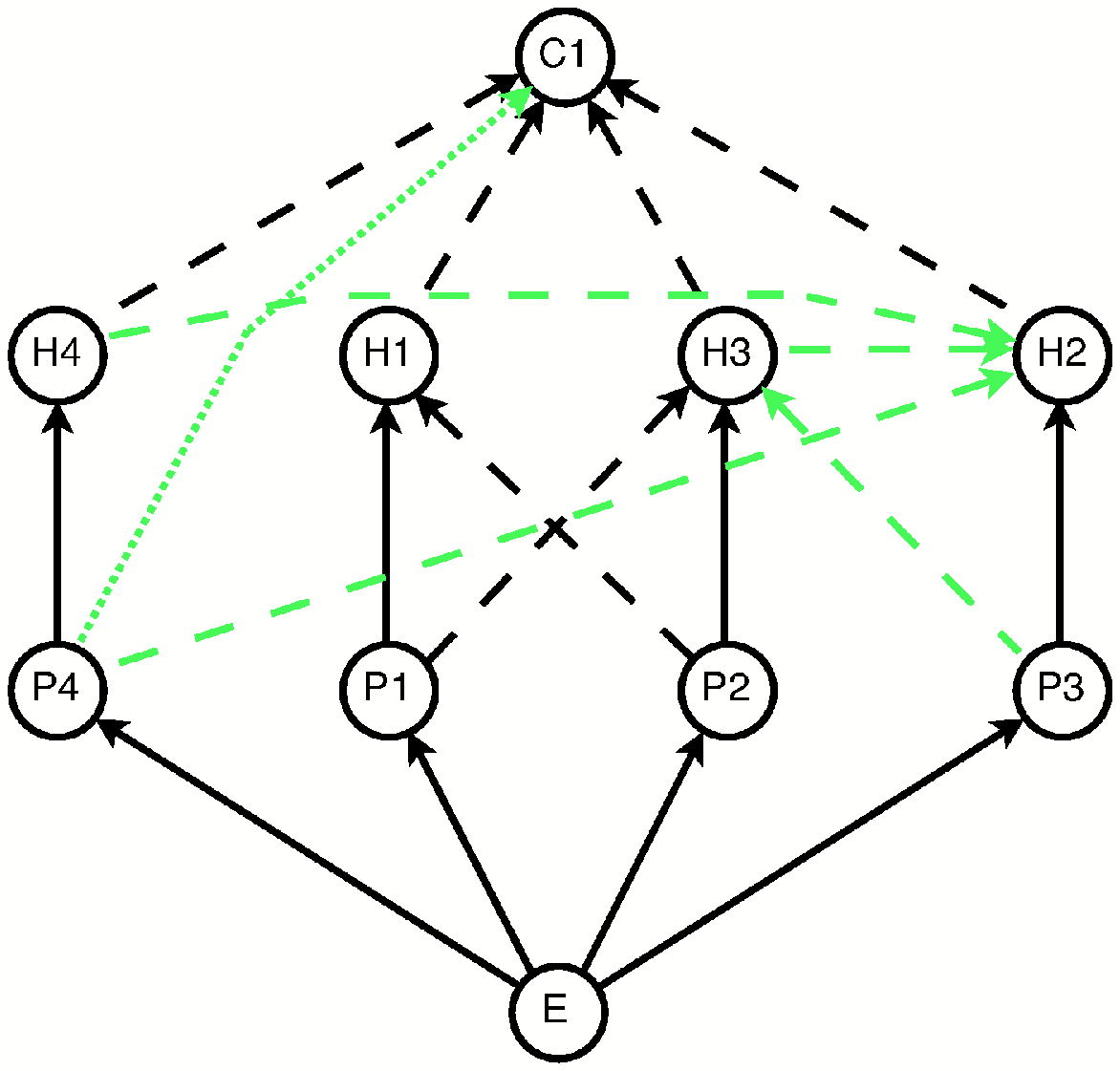}}\\ 
   
  \end{tabular}
  \caption[]{
    The food webs used in the comparison of time-series
    between the stochastic and parent population dynamics. In all
    webs, solid lines remain over all scalings. Long, medium and short
    dashed lines are those which are intermittent at $\Omega=10$,
    $\Omega=30$ and $\Omega=100$, respectively. The green
    arrows show intermittent links which exist only in the
    stochastic dynamics and are not represented in the parent model at
    all. The species designations are given in ascending order of
    population in the parent model.
  }
  \label{Tab:Webs}
\end{table}
A simple feature of Webworld, which arises from the use of a
ratio-dependent functional response, is that the dynamics of the
parent model are completely unaltered if the basal resources, $R$, the
population at which species go extinct and the population of each
species are all scaled by the same factor.  Equations (\ref{Eq:NDot})
and (\ref{Eq:g}) show that, once scaling is taken into account, the
population dynamics of the system is unchanged.  In the parent model
it is therefore reasonable to set the scale by assigning the
extinction threshold to a population of unity, but this does not
necessarily correspond to one individual of the stochastic model.
When comparing simple population dynamics with the parent model, we
therefore adopt a scaling factor, $\Omega$, by which populations from
the parent model are multiplied to obtain starting populations for the
stochastic model.  The results presented in
Figs.~\ref{Fig:EnsembleTimeSeries300}, \ref{Fig:EnsembleScaling},
\ref{Fig:ExampleTimeSeries300}, \ref{Fig:AnticorrelatedTimeSeries} and
\ref{Fig:PopulationVersusMu} are
divided by the same scaling factor to assist comparison.  The largest
value of scaling adopted is $\Omega=300$, which is a limitation of
computational resources.  The smallest value, $\Omega=10$, is a
consequence of the large fraction of simulation runs in which at least
one species goes extinct within the first one hundred units of time,
and suggests that the minimum viable population on this time scale is
of the order ten.  When considering island or reduction results, it is
no longer appropriate to refer to a scaling, since species are removed
from the community by extinction by fluctuations about an unknown
typical minimum population.  Instead,
Figs.~\ref{Fig:IslandLevelsVersusSpecies} and
\ref{Fig:IslandTopVersusSpecies} use the number of
species present as a proxy for food web size when comparing other food
web attributes between the parent and stochastic Webworld models.

Having introduced variant II, in which the foraging strategy adopted
by a species is not the evolutionarily stable strategy (ESS) but
rather a collective strategy based on each individual feeding on a
single prey species, it is important to assess the effect of choosing
different values of the mutation probability, $\mu$.  The demographic
extinction of certain strategies is inevitable if inheritance is
important, and even when $\mu=1.0$ individuals retain a single
strategy for their entire life-span, restricting the ability of the
strategy of the species as a whole to adapt to the contemporary ESS.
We therefore consider the effects of five values of $\mu$, spaced
logarithmically in the range $0.01\le\mu\le1.0$, to which we add
variant I\emph{b} as a comparable variant in which the ESS is followed at all
times.

The following three sections concentrate on three phenomena by which
the parent and stochastic models can be compared.  In
Section~\ref{Sec:Recovery} we examine how the parent model differs
from the ensemble average of results from the stochastic model.  In
Section~\ref{Sec:Fluctuations} we examine the nature of the
fluctuations about the ensemble average seen in any given realisation
of the stochastic model, and use these insights to understand some
attributes of the webs persistent under the stochastic dynamics.  In
Section~\ref{Sec:StrategyEffects} we consider in detail the effects of
using variant II\emph{b}, in which links between species are formed and
broken through the population demographics.

\subsection{Recovery of the parent model}
\label{Sec:Recovery}
The ensemble-average time series from the stochastic dynamics in
Island A at $\Omega=300$ and $\mu=0.1$ is reproduced in
Fig.~\ref{Fig:EnsembleTimeSeries300}. Apart from the least populous plant, the
ensemble average population of each species reaches a fixed point by
the mid-point of the time-series which differs by less than 3\% from
that of the parent model, which is the initial condition.  The
discrepancy relates to the extent to which variant II\emph{b} is able to sustain those
weak links which have been shown by \citet{mcc98} to have an important
influence on food web structure.  No extinctions were observed for
this value of $\Omega$, and excellent agreement with the parent model
is achieved for modest population sizes - each species present has a
population of less than ten thousand individuals.

For smaller values of $\Omega$,  the influence of demographic
stochasticity is more pronounced.  Not only do weak links become more
susceptible to disruption, but species as a whole are more vulnerable
to extinction.  In order to understand a homogeneous data set we
average over only those simulation runs in which extinction did not
occur.  Ensemble averages of the population dynamics, corresponding to
Fig.~\ref{Fig:EnsembleTimeSeries300}, are shown in
Fig.~\ref{Fig:EnsembleScaling} for the three species which
showed significant deviation from the steady-state population found in
the parent model.  The probable cause of the deviations from the
parent model result is the change in the food web structure as weak
links are removed, and is discussed in detail in
Section~\ref{Sec:StrategyEffects}.  The qualitative effect is that as
smaller values of $\Omega$ are considered, the parent model fails to
reflect even the ensemble average of the population dynamics.  The
important influence of weak links on the food web structure is
intimately connected with the population dynamics, since a weak link
implies an event of rare occurrence whether or not the species involved
are of small population.  The ensemble average for small $\Omega$ can
be recovered in the parent model by specific selection of which links
need to be incorporated, but this process cannot be properly
understood without first assessing the results of the stochastic
model.

\begin{figure}
  \includegraphics[width=0.45\textwidth]{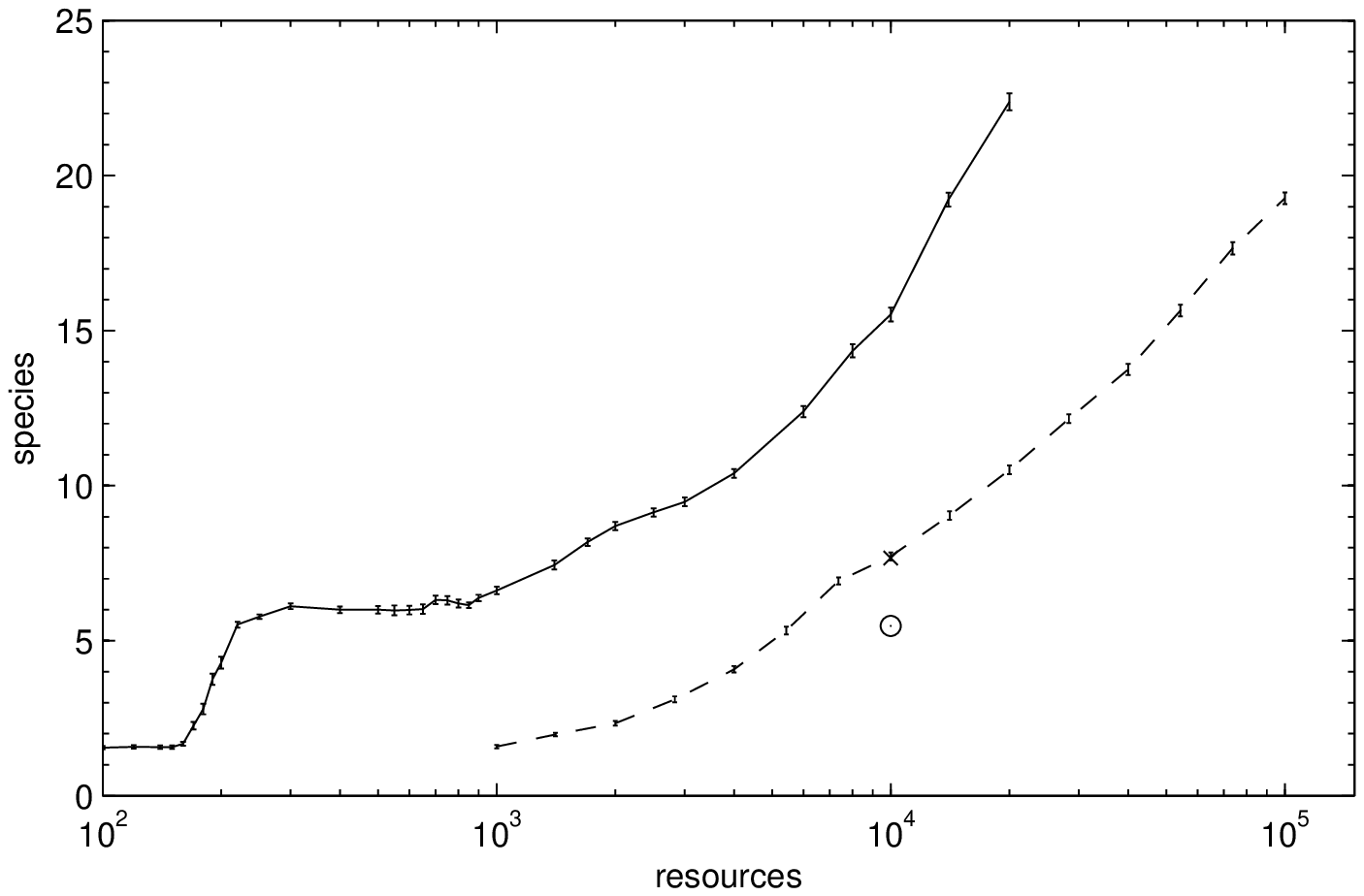}
  \caption{
    The mean number of species for an ensemble of one hundred
    islands, against resources.  Solid line; parent model.  Dashed
    line; variant II\emph{b} with $\mu=0.1$.  Cross; variant II\emph{a} with $\mu=0.1$.
    Circle; variant I\emph{b}.
  }
  \label{Fig:IslandSpeciesVersusR}
\end{figure}
\begin{figure}
  \includegraphics[width=0.45\textwidth]{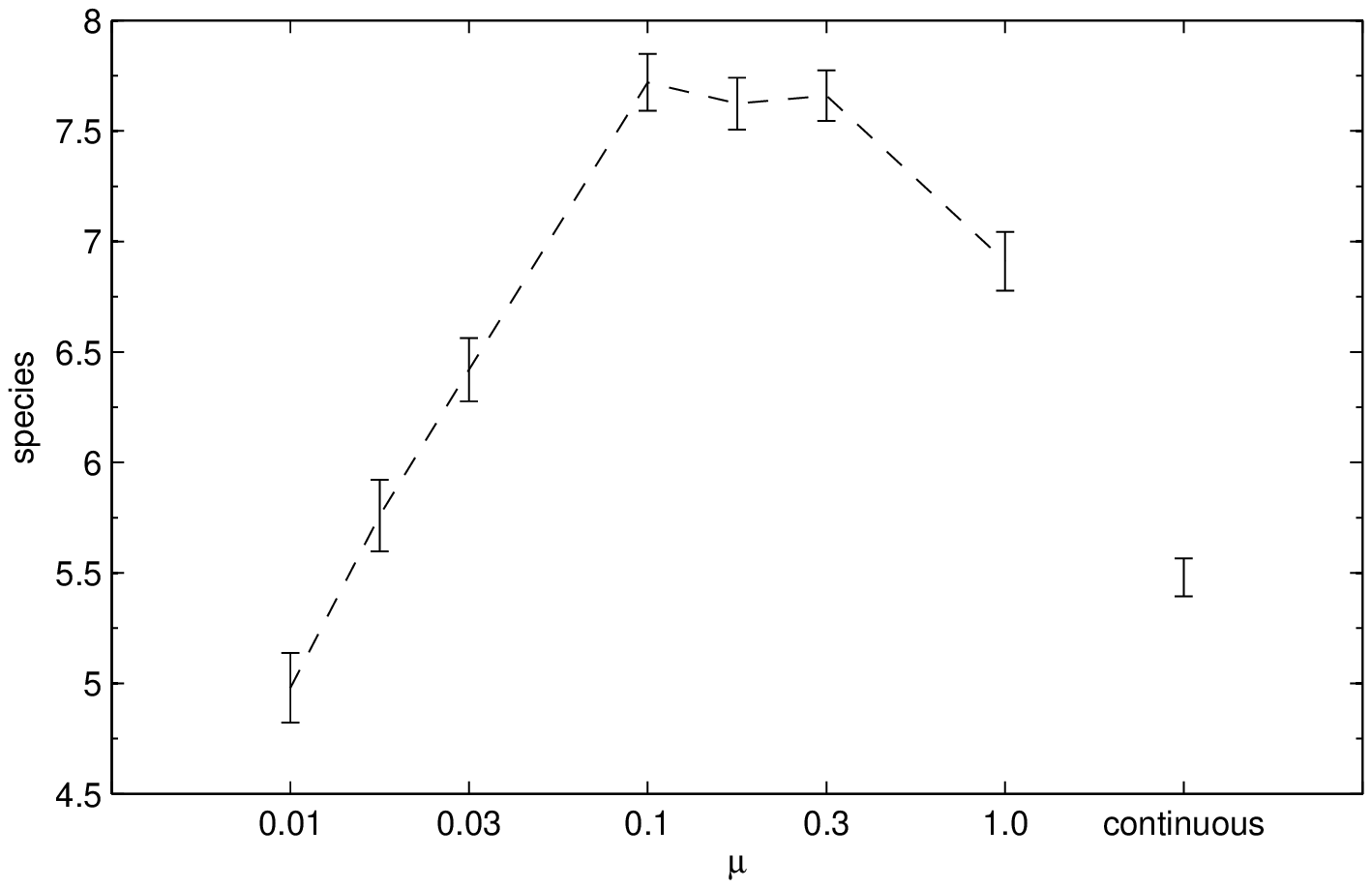}
  \caption{
    The mean number of species for an ensemble of one hundred islands
    against mutation probability $\mu$, for $R=10^4$.  The dashed line
    connects points belonging to variant II\emph{b}; the final, disconnected
    point corresponds to variant I\emph{b} and hence is not part of the same
    interpolation.
  }
  \label{Fig:IslandSpeciesVersusMu}
\end{figure}

The corresponding result for the island model is shown in
Fig.~\ref{Fig:IslandSpeciesVersusR}, in which the number of species
in a mature island community is shown for both the parent model and
variant II\emph{b} of the stochastic model.  The first point of agreement
with the population dynamics results is that there exists an
approximate factor of ten between the resources necessary to produce
comparable results in the two models; if the curve for the stochastic
dynamics is shifted a decade to the left, it approximately overlays
the parent model results.  In particular it is notable that complex
communities are still readily formed by immigration under stochastic
population dynamics, the largest communities constructed being
limited by computational resources rather than intrinsic properties of
the model.  Noting that these results correspond to the Type IV
species-area curve of \citet{sch03}, which relates to the number of
species in distinct islands and is therefore not necessarily
monotonic, there is a good overall agreement about the increase in
species number with resource availability, which we equate to the
physical extent of a real ecosystem.  The most significant deviation
corresponds to communities at the smallest values of resources, where
there is a step-like transition in the case of the parent model from
systems with one or two species to those with several.  This
phenomenon was discussed in \citet{Islands}, and is caused by the
weakening of niche-sharing restrictions on plants when one or more
herbivores enter the system.  In the case of the stochastic population
dynamics there is no noticeable step, and examination of numerous
example food webs indicates that even when herbivores become viable,
it remains difficult to sustain multiple plant species.  The
underlying cause of the continued exclusion under stochastic dynamics
is explored in Section~\ref{Sec:Fluctuations}.  The circle in
Fig.~\ref{Fig:IslandSpeciesVersusR} marks the number of species
present when variant I\emph{b} is used rather than variant II\emph{b}, and there is
consequently no selection against weak links.  The result is that
fewer species are able to co-exist, a consequence of the improved
ability of herbivores to exploit relatively rare plant species.
Fig.~\ref{Fig:IslandSpeciesVersusMu} shows how the number of species
simultaneously sustained for this value of resources, $R=10^4$, varies
with the ability of species to adopt new strategies though changing
mutation, $\mu$.  The point corresponding to variant I\emph{b} does not
properly belong on the same scale as different values of $\mu$, but
conceptually corresponds to a situation of very large mutation rate
and is therefore plotted to the right of the figure.  It is notable
that there exists a peak in the number of species, where small values
of $\mu$ do not allow sufficiently diverse strategies to maximise the
species diversity, but large values of $\mu$ also reduce diversity
through the ease of resource exploitation, indicating a positive
relationship between specialisation and diversity for this region of
parameter space.

\begin{figure}
  \includegraphics[width=0.45\textwidth]{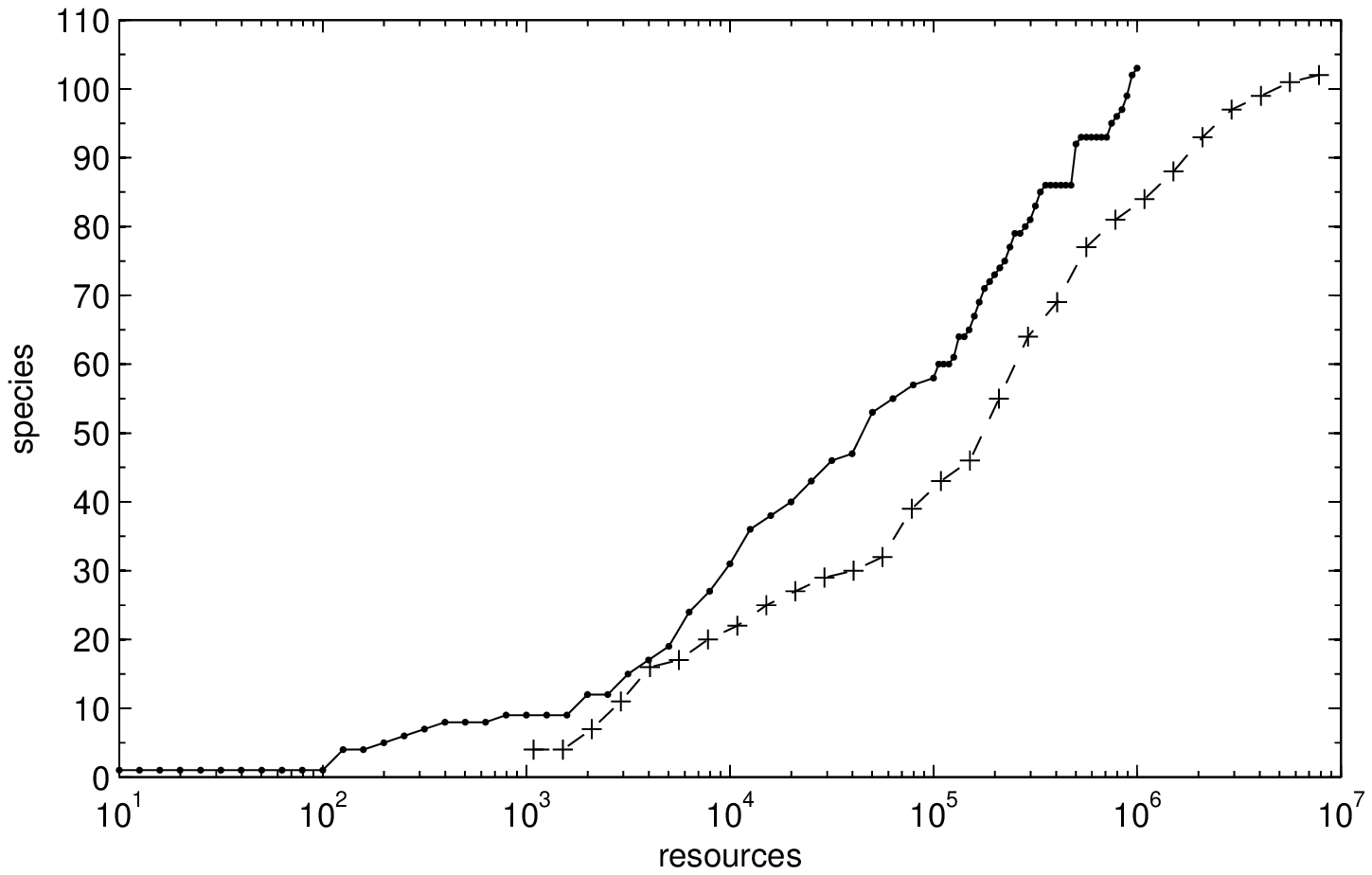}
  \caption{
    The number of species remaining as resources are reduced, against
    resources.  Solid line; parent model with slow reduction.  Dashed
    line; variant II\emph{b} with $\mu=0.1$, reducing resources by 8\% in unit
    time, compared with $d=1$.
  }
  \label{Fig:ReductionSpeciesVersusR}
\end{figure}

A further check on the ability of the stochastic population dynamics
to support complex food webs is supplied by the reduction scenario.
Fig.~\ref{Fig:ReductionSpeciesVersusR} shows the path that a single
stochastic realisation of the reduction of resources follows, along
with the unique path followed by the parent model.  Similarly to the
island results, there is approximately a factor of ten in resources
between the two curves, although this is partially obscured by the
rapid rate at which it was necessary to constrict resources for the
stochastic model.  Whereas the population dynamics of the parent model
can be run such that resources are reduced at an effectively
infinitesimal rate, computational requirements and practical
considerations require the stochastic model to be run at a definite
reduction rate.  Computing the population dynamics of any community
under stochastic population dynamics for an infinite length of time
without speciation will result in the extinction of every species
through fluctuations, so there does not exist an asymptotic curve for
the stochastic model.  Furthermore, the slowest rate of restriction
for which results could be simulated corresponds to an 8\% reduction
in resources per unit time period, which is rapid given that the
expected lifetime of an individual is also unit time.  Species are
therefore expected to persist to smaller values of $R$ under these
circumstances than would be true for a slower reduction rate, and the
curve is somewhat smoothed by the stochastic delay introduced between
the time a species becomes unviable and the moment at which it becomes extinct.
Allowing for this effect, there is good agreement between the two
curves, and the stochastic model is shown to be successful at allowing
large, complex communities to persist.

\subsection{Fluctuations and intra-trophic neutrality}
\label{Sec:Fluctuations}
\begin{figure}
  \includegraphics[width=0.45\textwidth]{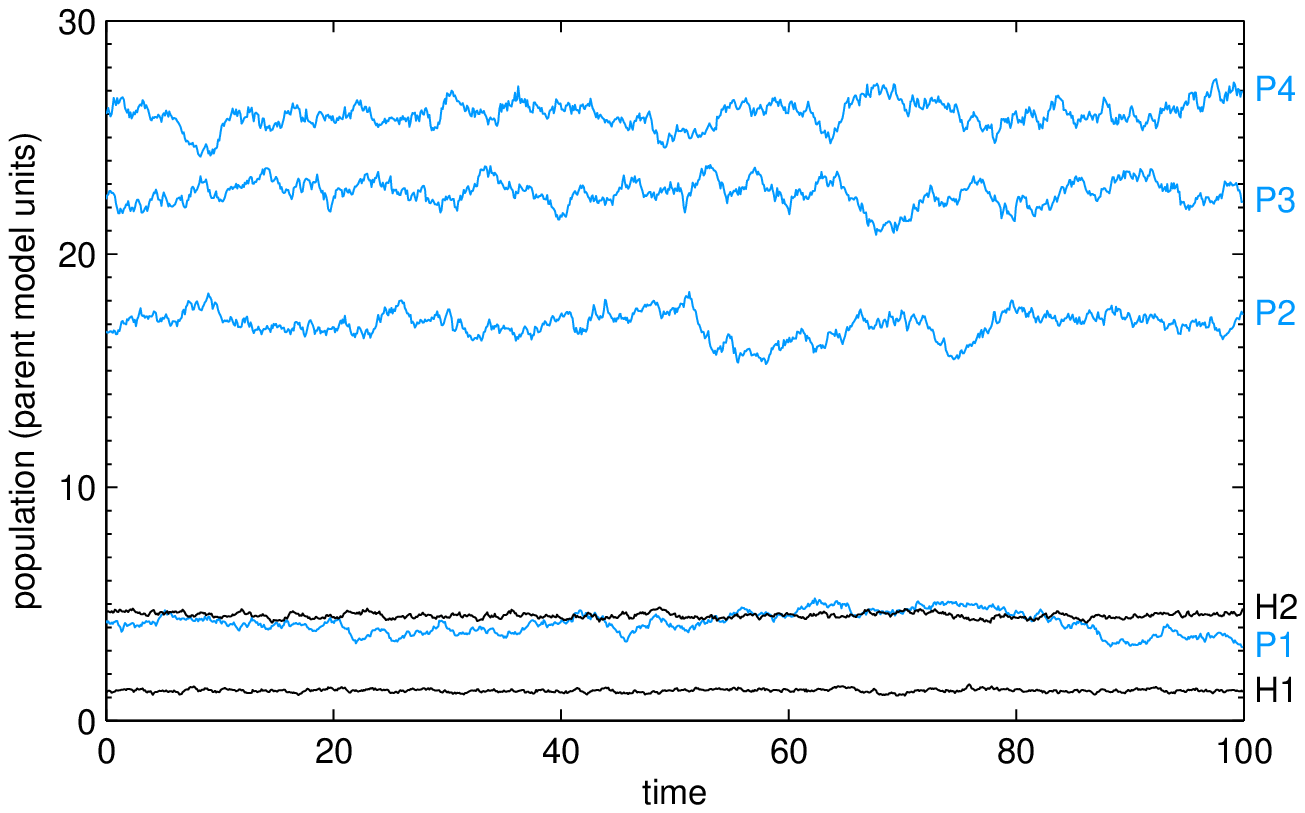}
  \caption[]{
    The time series of one realisation of the
    stochastic population dynamics for Island A, using variant II\emph{b} with
    $\mu=0.1$ and scaling $\Omega=300$.  Blue lines; plant
    species.  Black lines; herbivores.
  }
  \label{Fig:ExampleTimeSeries300}
\end{figure}

\begin{figure}
  \includegraphics[width=0.45\textwidth]{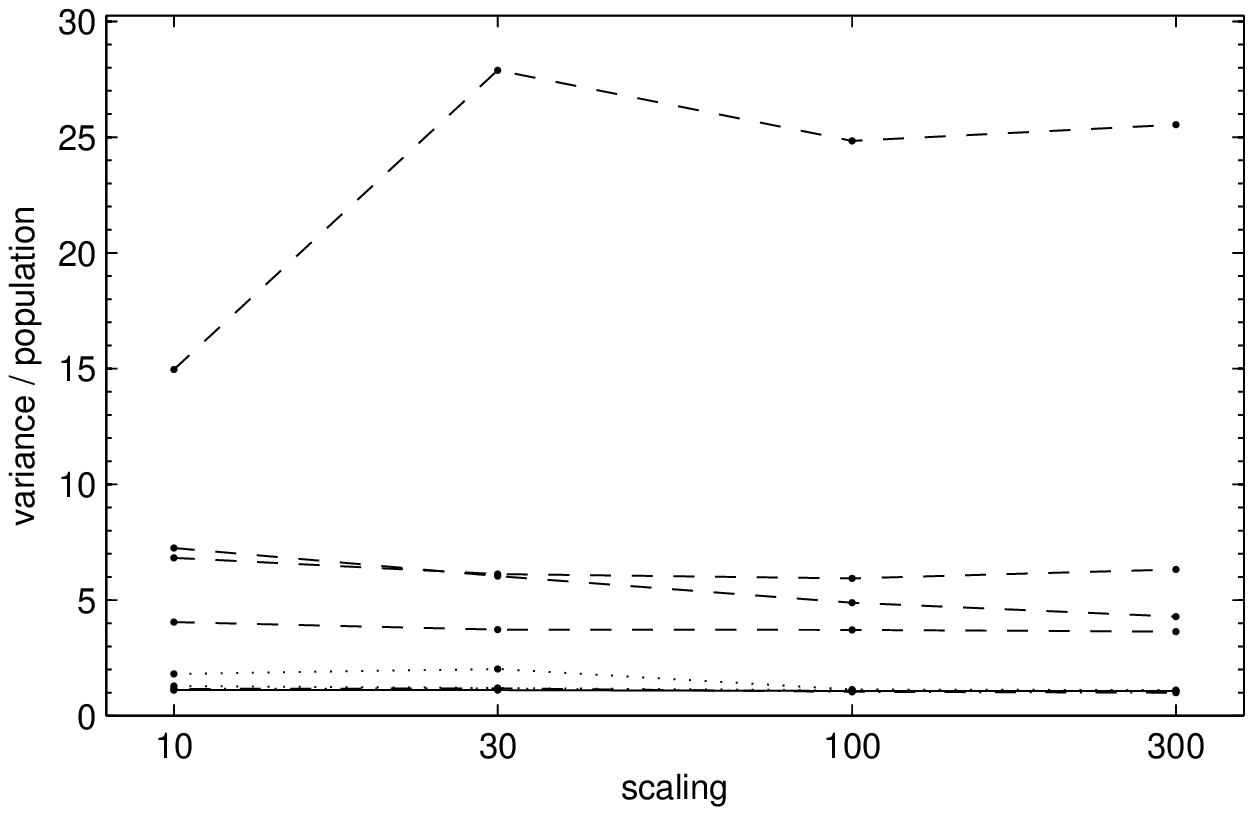}
  \caption[]{ 
    The ratio of the variance of population time series
    to the mean population at various values of scaling,
    $\Omega$. Dashed lines; plant species.  Dotted lines; herbivore
    species.  Solid line; total plant population.  Dash-dotted line;
    total herbivore population.
  }
  \label{Fig:FluctuationScaling}
\end{figure}

One way in which the stochastic model differs from the parent model is
that fluctuations in population sizes exist even when the ensemble
average populations are constant.  Extinction occurs when the
fluctuating population of a species drops instantaneously to zero, and
hence the relative magnitude of fluctuations is an important
contributing factor to food web structure.  In this section we examine
the variance of those fluctuations, and the extent to which the
fluctuations in different species are not independent.
Fig.~\ref{Fig:ExampleTimeSeries300} shows that, even for scaling
$\Omega=300$, these fluctuations are of significant amplitude, and it
is notable that despite the similarity in population size, plant P1
exhibits significantly larger fluctuations than herbivore H2.
Fig.~\ref{Fig:FluctuationScaling} shows the variance of each of the
species averaged across an ensemble for each of the scaling factors,
$\Omega$, examined.  In every case the variance for plant species is
larger than that for the herbivores, the more populous of which has
a variance nearly equal to its mean population for all values of 
$\Omega$ used.  For the larger values of $\Omega$ the variance
is approximately proportional to the population for all species, and hence the lines
in Fig.~\ref{Fig:FluctuationScaling} are approximately horizontal.
This indicates that as the scaling factor increases, the standard
deviation of population size increases only as the square root of the
population.  The ability to
neglect fluctuations for very large population sizes is an assumption
implicit in the parent model, and these results demonstrate that the
fluctuations do indeed become negligible.

\begin{figure}
  \includegraphics[width=0.45\textwidth]{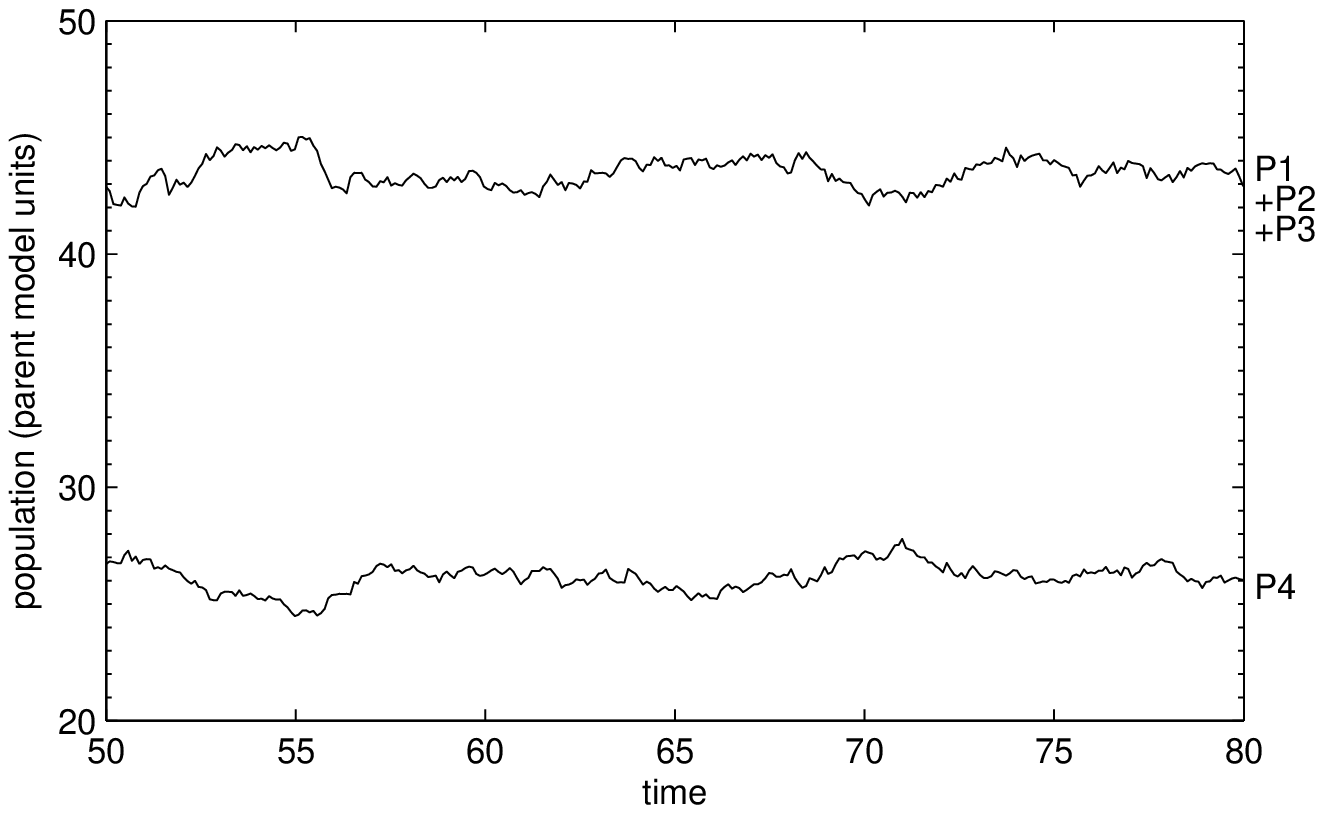}
  \caption{
    An example of anti-correlations between
    populations of basal species, taken from Island A using variant
    II\emph{b} with $\mu=0.1$, scaling $\Omega=300$. 
    The lower line marks the population of the most populous plant, P4.
    The upper line marks the total population of the other three plants.
  }
  \label{Fig:AnticorrelatedTimeSeries}
\end{figure}

If time series subject to independent noise are added, the variance of
the summed time series is equal to the sum of the variance of each
time series separately.  The relative variance would therefore lie between the values
calculated for each time series, yet
Fig.~\ref{Fig:FluctuationScaling} clearly shows that this is not the
case for the stochastic Webworld results.  Indeed, the variance for
the sum of the plant populations is nearly equal to the total
population, as is also the case for the sum of herbivore populations.
This indicates that the fluctuations in the plant populations must be
anti-correlated.  In Fig.~\ref{Fig:AnticorrelatedTimeSeries} a time
series for plant P4 is shown along with that for the sum of the
other three plants.  It can be seen that significant anti-correlation
does exist, which is especially obvious for the large fluctuations around time
55 and 70.  To quantify this relationship we calculate the correlation
coefficient, $r^2$, between time series \citep{hog94}. It was found that an
increase in $r^2$ could be found by allowing a offset between the time
series of different species, and Table~\ref{Tab:Correlations} records
the maximum value of $r^2$ and the corresponding time offset, along
with the gradient of the best-fitting regression line.  The
results show that for Island A two pairs of plants exist with 
relatively strong correlation; species P1 correlating
strongly with P3, and P2
with P4.  As shown in Table~\ref{Tab:Webs}, these pairs form trophic species in the
full sense that they share both food source and predators, although P1
is also weakly predated by H2. These trophic species are
strongly correlated with each other, and are more strongly correlated
their respective herbivores than is any individual plant species. 
The optimal time offset for any pair of species within the same
trophic level is near zero, while the best correlation between the
trophic plants and herbivores is found with the herbivore time series
lagging by up to approximately unit time, suggesting that changes in
the herbivore population can, in part, be attributed to fluctuations in the abundance of
their food source. 
The gradient of the regression line between each pair of plant species
is close to $-1$. This is a
consequence of the strong competition between species which share a
common food source. Although the total population across all plant
species is relatively well-constrained, any one plant species can
increase in population at the expense of it competitors.  Meanwhile
there is a positive correlation between the population of each
herbivore and that of the trophic
species upon which it predominantly feeds. With the opposite gradient,
there is a consequent correlation between herbivore population and
the competitors of their prey.

A similar analysis was carried out with Island B,
of which the results for the sub-web comprising P1, P2, H1 and H3 are
particularly notable and are appended to
Table~\ref{Tab:Correlations}. Two major differences between this
sub-web and Island A are the presence of stronger cross-links
and the presence of the carnivore, making the combination of H1 and H3
a trophic species. As well as the correlation between plants,
there is a significant correlation between herbivores, also with
near-zero offset.
However, although the optimal
offset between herbivore and plant time series is of similar magnitude
to those seen Island A, in Island B it is the herbivore populations
which lead. 
The gradient in Table~\ref{Tab:Correlations} is equal to the number of
individuals of the plant species necessary to support a single extra
herbivore. This number is greater in Island B due to the additional
losses to predation by the carnivore, whose inclusion
may be responsible for altering the relative strength of top-down and
bottom-up effects, reflected in the change from lagging to leading
herbivore fluctuations.

\begin{table}
\center  
\begin{tabular}{c|cc|ccc}
    Island&\parbox{1.3cm}{$x$}&\parbox{1.3cm}{$y$}&\parbox{1.3cm}{$r^2$}&\parbox{1.3cm}{$\Delta t$}&\parbox{1.3cm}{gradient} \\
    \hline
    \multirow{12}{*}{A} & P1 & P2  & 0.04 & 0.0 & -1.0 \\
    & P1 & P3 & 0.16 & -0.2 & -1.0 \\
    & P1 & P4 & 0.04 & -0.2 & -0.99 \\
    & P2 & P3 & 0.08 & -0.2 & -0.98 \\
    & P2 & P4 & 0.16 & -0.1 & -0.97 \\ 
    & P3 & P4 & 0.02 & -0.0 & -0.94 \\ 
    & P1+P3 & P2+P4  & 0.47 & 0.0 & -1.0 \\
    & H1 & P1+P3  & 0.10 & -0.5 & 3.8 \\
    & H1 & P2+P4  & 0.083 & -0.7 & -3.8 \\
    & H2 & P1+P3  & 0.031 & -1.1 & -2.4 \\
    & H2 & P2+P4  & 0.059 & -0.7 & 2.5 \\ 
    & H1 & H2 & 0.030 & -0.3 & -1.38 \\ \hline
    \multirow{6}{*}{B} & P1 & P2  & 0.60 & -0.1 & -1.0 \\
    & H1 & P1  & 0.69 & 1.3 & 11 \\
    & H1 & P2  & 0.66 & 1.3 & -11 \\
    & H3 & P1  & 0.59 & 1.7 & -9.8 \\
    & H3 & P2  & 0.71 & 1.0 & 10 \\ 
    & H1 & H3 & 0.60 & 0.1 & -1.0
  \end{tabular}
  \caption[]{   
    Selected data from the analysis of
    correlations in both Islands A and B. The $y$-time series is
    offset by $\Delta t$ with respect to the $x$-series to maximise the
    correlation coefficient, $r^2$. The gradient shown is that of the
    best-fitting regression line at $\Delta t$. Note that the typical
    population of species in Island B is greater than in Island A,
    contributing to larger values for $r^2$.
  }
  \label{Tab:Correlations}
\end{table}
In the parent model, the co-existence of multiple species within a
trophic species is allowed because inter-specific competition is less
that intra-specific competition, encoded in the competition factor,
$\alpha$.  This can be interpreted as reflecting differences in the
feeding habits of competitor species which are not explicitly modelled.
Under stochastic
population dynamics an additional effect is observed in which
demographic fluctuations in the populations of the niche-sharing
species are larger than those of species which do not share a
niche.  By extension we expect that an increasing overlap of common
predators and prey between two species corresponds to an
increasing amplitude of fluctuations in population, while the total
population of the niche (trophic species) remains relatively
steady.  As inter-specific competition increases, and
$\alpha\rightarrow1$, both the parent and stochastic models would
predict the extinction of the less suitable taxonomic species within
trophic species. 

A consequence of the large fluctuations in the population of
trophically similar species in
the stochastic model is a stronger exclusion of niche-sharing
species. In turn, fewer plants are able to co-exist in island
communities.  This is reflected in the reduced number of 
species present in stochastic islands relative to parent-model islands,
as can be seen in Fig.~\ref{Fig:IslandSpeciesVersusR}.  A further
effect, shown by Fig.~\ref{Fig:IslandLevelsVersusSpecies}, is that the average number
of trophic levels is higher in the case of the stochastic model
than for the parent model when comparing webs with an equal number of
species and no carnivores (approximately, $S<10$).  Since webs 
containing
multiple plants and no herbivores are less stable under stochastic
dynamics, it follows that for a given number of species there will be
more webs with at least one herbivore.  

\subsection{Strategy extinction and web restructuring}
\label{Sec:StrategyEffects}
Having introduced demographic effects into foraging strategy through
variant II of the stochastic model, in which each individual feeds on
a single prey species for its entire life-span, species
become susceptible to
`forgetting' feeding links if the typical number of individuals
following that strategy is small.  As the scaling factor, $\Omega$, is
reduced, species become increasingly susceptible to this effect, and
departures from the food web structure of the parent model are
expected to increase in magnitude.  The diagrams of the Island A and
B food webs shown in Table~\ref{Tab:Webs} illustrate how weaker feeding links
are successively lost as $\Omega$ is decreased from 300, when the webs
are essentially equivalent to those in the parent model, through
$\Omega=100$, 30, and 10.  Once a feeding link is removed from the
model, the prey species typically increases in population and the
predator decreases, since 
it is receiving income from fewer prey individuals.
The context within the food web structure complicates this picture,
since the removal of competition strengthens the ability of other
predators to exploit a species.  For example, if the link between P1
and H2 in Island A is lost, H1 does not suffer inter-specific
competition when feeding on P1, and is therefore able to increase in
population.  This reduces the beneficial effect to P1 of this
link being lost. 
A second example of the indirect consequences of the loss of weak
links is provided by the cessation of feeding by H2 on H1.  When this
occurs H1 increases in population, increasing the predation pressure
on plants P1 and P3, whose populations decrease.
The propagation of complex indirect effects through
the food web can be understood in basic terms by investigation of the
proximal effects of link removal, but the net effects on
the community cannot be 
predicted any more simply than by directly simulating
the population dynamics, due to the complex nature of
competition and predation relationships.

\begin{figure}
  \includegraphics[width=0.45\textwidth]{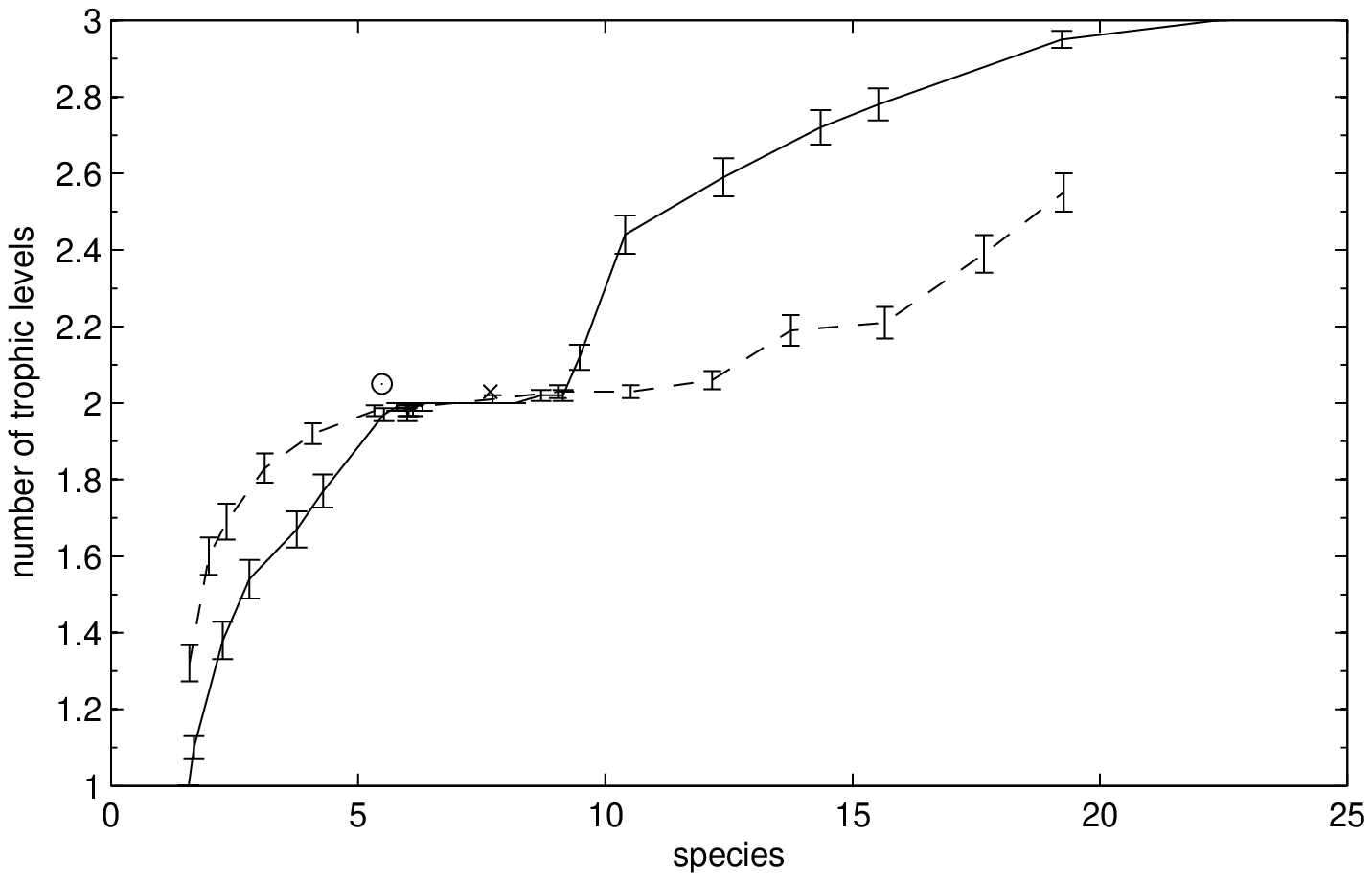}
  \caption{
    The mean number of trophic levels against the mean
    number of species present, for an ensemble of one hundred islands
    at each point.  Solid line; parent model.  Dashed
    line; variant II\emph{b} with $\mu=0.1$.  Cross; variant II\emph{a} with $\mu=0.1$.
    Circle; variant I\emph{b}.
  }
  \label{Fig:IslandLevelsVersusSpecies}
\end{figure}
In addition to the loss of weak links, some links appear in the
stochastic model which would be predicted from the parent model to
correspond to less than a single individual.  For small values of
$\Omega$ the food web structure has been changed by the deletion
of other links, with the result that the populations of species differ from their
values in the parent model.  Under these circumstances links may
become viable for one of two reasons; either the prey species has
increased in abundance to the point at which it forms an exploitable
resource, or competition for that prey has decreased.  In general both
of these effects occur, and several such links are marked on the
diagram of Island B in Table~\ref{Tab:Webs}.  All these links are
intermittent - the predator species does not permanently have
individuals which follow the strategy, but that sub-population goes
extinct and is later re-established through the birth of a mutant
individual.  In the case of extremely simple food webs it can be seen
that there are population changes in the prey species depending on
whether or not it is being fed upon, but in even moderately complex food webs
such effects are within the demographic noise and hence cannot be
clearly demonstrated.

\begin{figure}
  \includegraphics[width=0.45\textwidth]{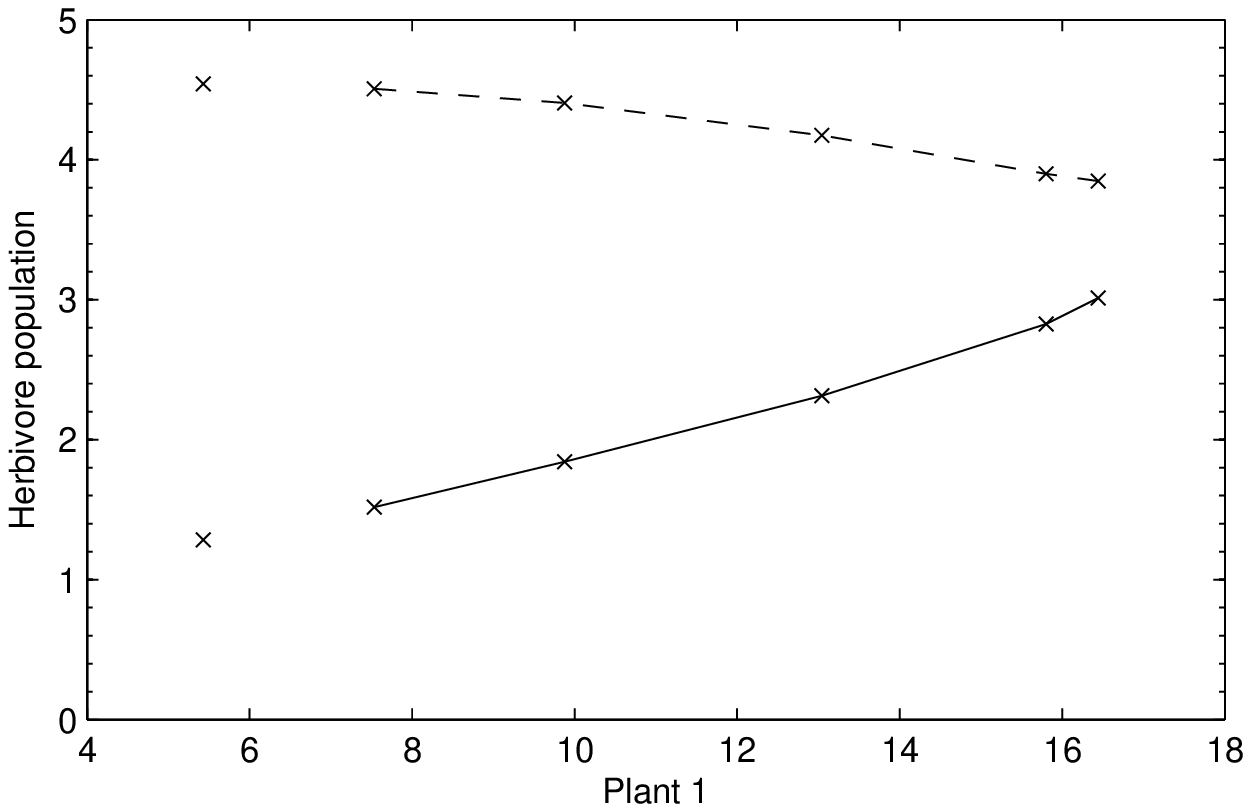}
  \caption{
    Parametric plot of the population of plant P1 and the two
    herbivores of Island A, using variant II\emph{b} with scaling $\Omega=50$.
    From right to left the points correspond to
    $\mu=0.01$, 0.03, 0.1, 0.3, and 1.0.  The disconnected points to
    the left correspond to variant I\emph{b}.
  }
  \label{Fig:PopulationVersusMu} 
\end{figure}
Fig.~\ref{Fig:PopulationVersusMu} demonstrates how the population of
plant P1 increases with decreasing $\mu$, which corresponds to an
increase in the expected time before an individual of species H2
mutates to re-populate the extinct foraging strategy.
Correspondingly, herbivore H2 is seen to decrease in population with
decreasing $\mu$, since this source of nutrients forms an increasingly
small part of its diet.  Herbivore H1, whose feeding link to P1 does
not significantly decrease over this range of $\mu$, experiences a
population increase correlated with the increasing abundance of its
prey.  Because H1 does not feed exclusively on this plant, its
population increase is smaller as a fraction of its total population
than the increase in P1, which becomes an increasing fraction of its
diet with increasing abundance.  This figure demonstrates specific
causes of both positive and negative correlations between populations,
illustrating the complex outcomes of the food web dynamics.

The large-$S$ part of Fig.~\ref{Fig:IslandLevelsVersusSpecies} shows
that the number of carnivores is less in the stochastic than the
parent model when considering food webs with equal numbers of species,
but that carnivores are not completely excluded from the food webs.
The reduction in their prevalence is a consequence of the removal of
weaker feeding links, since carnivores in the Webworld model tend to
have more diverse diets than do lower trophic levels.  As an example
of this, the single carnivore in Island B feeds on each of the four
herbivores, but no herbivore in the parent model (i.e.\ excluding shaded
links in Table~\ref{Tab:Webs}) feeds on more than two species.  Because the population
required to sustain links to four prey species is greater than that
necessary to feed on a single prey, the entry of carnivores into the
food web is delayed relative to the entry of the lower trophic levels,
and in particular it is possible to support a larger number of plant
and herbivore species before carnivorous strategies become viable.  As
a consequence, the curve for the stochastic model lies below that for
the parent model when considering trophic levels greater than two.
The absence of these feeding links also increases the fraction of
`top' (unpredated) species, as shown in
Fig.~\ref{Fig:IslandTopVersusSpecies}.  For a small total number of
species, $S$, the curves for the parent and stochastic models are
very similar as only plants and herbivores exist, and despite
differences in the proportion of species found in those two trophic
levels as discussed in Section~\ref{Sec:Fluctuations}.  The circle
marks the point corresponding to variant I\emph{b}, in which the presence of
feeding links is not affected by demographic effects, and which as a
consequence is within the region occupied by the parent model.  The
inset to Fig.~\ref{Fig:IslandTopVersusSpecies} shows the monotonic
decrease in the fraction of top species as $\mu$ is increased, and the
continuation of this trend into variant I\emph{b}.

\subsection{Refuge for sated individuals}
\label{Sec:Stupidity}
\begin{figure}
  \includegraphics[width=0.45\textwidth]{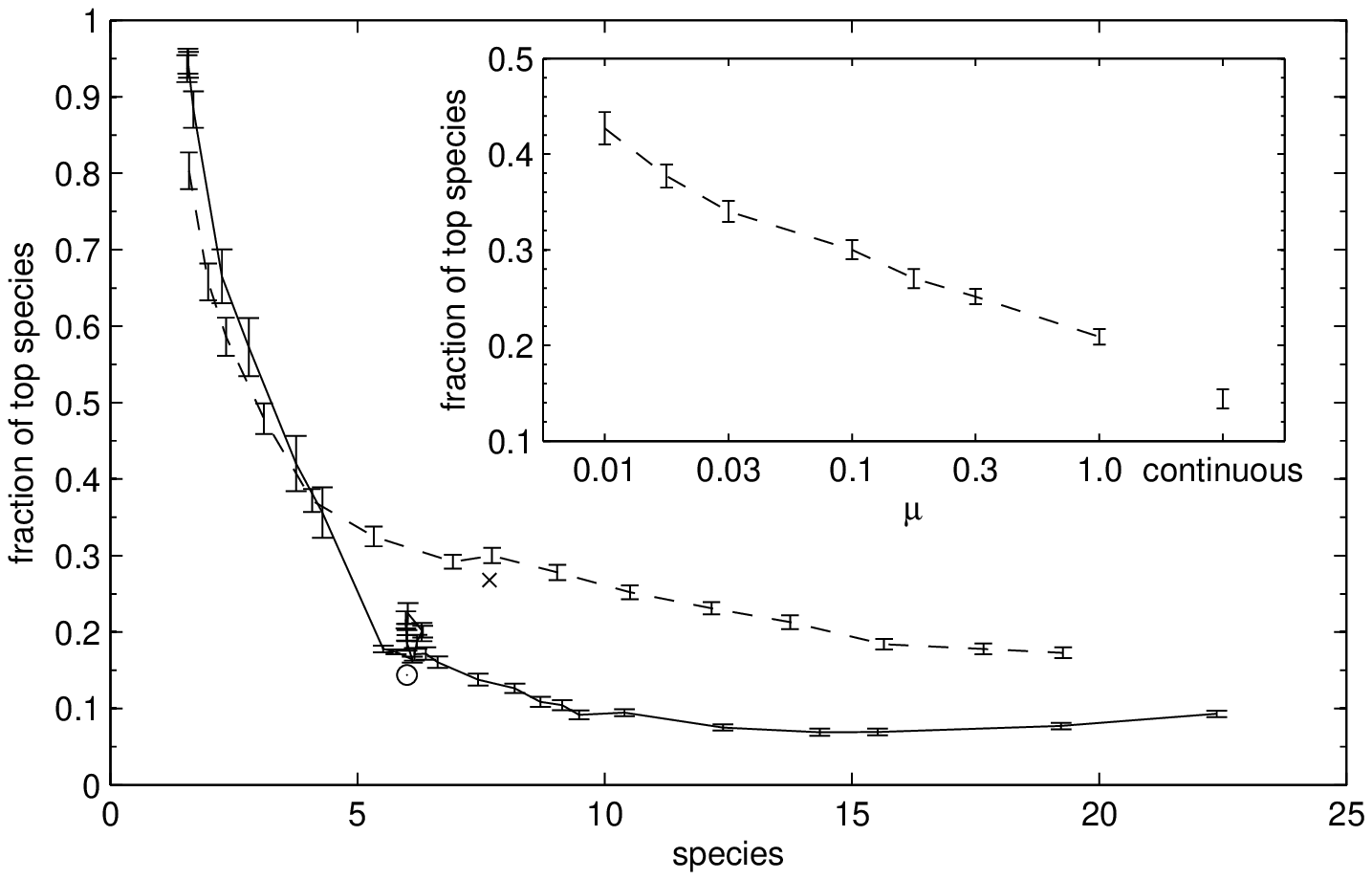}
  \caption{
    The mean fraction of unpredated species against the mean number of
    species present, for an ensemble of one hundred islands
    at each point.  Solid line; parent model.  Dashed line; variant
    II\emph{b} with $\mu=0.1$.  Cross; variant II\emph{a} with $\mu=0.1$.  Circle;
    variant I\emph{b}.  {\bf Inset:} The mean fraction of unpredated species
    against mutation probability $\mu$, for $R=10^4$.
  }
  \label{Fig:IslandTopVersusSpecies}
\end{figure}
The distinction between variants \emph{a} and \emph{b} of the model corresponds to
the ability of sated individuals, who cannot gain advantage from any
interaction, to avoid interaction with predators.  In variant \emph{b} sated
individuals are completely safe from predation, whereas in variant \emph{a}
they are as vulnerable as foraging individuals are.  The plausible
behaviour of a more detailed model is likely to lie between these two
extremes.  Table~\ref{Tab:Stupidity} records the ensemble average
population for each of the six species of Island A for variants II\emph{a}
and II\emph{b}, and variant I\emph{b} for comparison.  The standard deviations, also
shown, indicate the size of the population fluctuations.  It is
apparent that for each species the difference between variants II\emph{a} and
II\emph{b} is much smaller than the corresponding fluctuations, but the
differences between variants I and II are significant.  These findings
are reflected in the position of markers in
Figs.~\ref{Fig:IslandSpeciesVersusR},
\ref{Fig:IslandLevelsVersusSpecies} and
\ref{Fig:IslandTopVersusSpecies}.  In each case variant II\emph{a} produces
results essentially equivalent to variant II\emph{b}, while variant I\emph{b} shows
significant differences.  In the latter two figures, where the number
of species in an island food web is used to relate the size of
parent-model and stochastic-model webs, variant I\emph{b} produces results
similar to the parent model.  
If refuge has important consequences for
food web structure
%, which is probable given its prevalence as an
%ecological concept \citep{...}, 
then its effects are likely to be concealed
within the functional response used by Webworld.

\section{Conclusions}
The primary aim of this paper was to develop the Webworld model by
introducing an individual-based population dynamics.  The Webworld
model has been successful in constructing food webs with several
realistic properties, based on simple population dynamics and foraging
strategy optimisation.  By developing a model whose basic assumptions
correspond to a finer level of detail, the interactions between
individuals rather than species-level descriptions, it becomes
possible to explore the consequences of a range of behaviours enacted
at the individual-level.  For instance, at the level of the parent
model it is possible to test robustness to different forms of the
functional response, whose normal form is given by (\ref{Eq:g}), but
use of an individual-based model allows the more direct comparison of
different forms of the reaction rate equations which are assumed to
underlie species-level interactions.  In addition, it has been shown
that the stochastic model closely approximates the parent Webworld
model when population sizes are large, but that when considering
constructed food webs this is never an acceptable assumption, since
there always exist some species which would not be viable for a
significantly smaller ecosystem, and whose population demographics
have a profound influence on the community as a whole.

\begin{table}
  \begin{tabular}{l|cc|cc|cc}
    
    &\multicolumn{2}{|c|}{Variant II\emph{a}}&\multicolumn{2}{|c|}{Variant I\emph{b}}&
    \multicolumn{2}{|c}{Variant II\emph{b}} \\ 
    Species &
    $\mu$ & $\sigma$ &
    $\mu$ & $\sigma$ & 
    $\mu$ & $\sigma$ \\ \hline 
    P1 & \parbox{1.0cm}{12.78} & \parbox{1.0cm}{4.8}  &  \parbox{1.0cm}{5.9}  & \parbox{1.0cm}{2.7}  & \parbox{1.0cm}{13.10}  & \parbox{1.0cm}{4.5}  \\ 
    P2 & 15.49 & 3.3  & 16.8  & 2.9  & 15.64  & 3.2  \\ 
    P3 & 18.44 & 3.6  & 21.6  & 3.0  & 18.51  & 3.6  \\ 
    P4 & 24.07 & 3.1  & 25.6  & 3.0  & 24.24  & 3.1  \\ 
    H1 &  2.31 & 0.86 &  1.28 & 0.41 &  2.326 & 0.85 \\
    H2 &  4.22 & 0.82 &  4.60 & 0.72 &  4.164 & 0.81 \\ 
    
  \end{tabular}
  \caption[]{ 
    A comparison of stationary behaviour between variant II\emph{a}, in which
    all individuals are vulnerable to predation, and variant II\emph{b}, in which only
    hungry individuals are vulnerable, is shown for Island A. The scaling used 
    here is $\Omega=10$.  $\mu$; mean.  $\sigma$; standard deviation.
  }
  \label{Tab:Stupidity}
\end{table}
A secondary aim in developing an individual-based version of the
Webworld model was to establish a point on the way to an agent-based
model.  At the level of the individual, it becomes possible to model
ecologically important processes such as predator-evasion, foraging
choice through individual access to observations about prey
distribution, and to include the effects of life-history decisions and
mating strategy.  Each of these can be expected to have effects on the
population dynamics of the species and hence on the structure of the
ecosystem, but it is not practically possible to create a simulation
model which takes individuals with detailed behaviour as a starting
point and from that constructs a food web structure.  Instead we have
adopted the strategy of taking a model known to produce realistic food
webs, within which framework we can examine at various scales the
influences that differences in the details of one level of abstraction
have on the outcome of a broader picture.  If a bridge can be built
linking choice at the level of the individual at one end to ecosystem
behaviour at the other, it will be possible to explore the
consequences of the former on the latter despite the practical
impossibility of computing the whole of an ecosystem at a sufficient
level of detail to resolve detailed variation between individuals.

Some of the important results to come out of the preliminary work on
the stochastic model performed in this paper relate to the effects of
demographic stochasticity in different food web positions.  In
particular it has been shown that interactions between species
significantly alter the magnitude of demographic fluctuations, and
that taxonomic species forming part of a larger trophic species are
more susceptible to these fluctuations than are species who share
neither predators nor prey.  The stochastic model broadly reproduces
the results of the parent Webworld model in terms both of population
dynamics and the construction of communities through immigration,
although it is necessarily the case that extinction in the stochastic
model occurs through fluctuations occasionally reducing a population
to zero, rather than the mean population reaching a certain threshold
as is the case in the parent model.  Webs constructed in the parent
model are robust to the imposition of the stochastic population
dynamics, and a future line of enquiry would be to establish whether
this is due to the degree to which the two sets of population dynamics
are equivalent, or whether the method of constructing food webs
through an evolutionary process of species addition selects for
robustness to changes in relative species abundance, and hence
provides stability against the demographic fluctuations.  

\section*{Acknowledgements}
We wish to thank EPSRC (UK) for funding; CRP under grant number
GR/T11784 and RPB under a postgraduate grant.

\end{document}